\documentclass[apjl]{emulateapj}
\usepackage{graphicx,natbib,longtable}
\usepackage{enumerate}
\begin{document}

\title{A Deep Search for Prompt Radio Emission from Thermonuclear Supernovae with the Very Large Array}

\author{Laura~Chomiuk\altaffilmark{1}$^{,}$\altaffilmark{11}, 
Alicia~M.~Soderberg\altaffilmark{2}, 
Roger~A.~Chevalier\altaffilmark{3}, 
Seth~Bruzewski\altaffilmark{1},
Ryan~J.~Foley\altaffilmark{4,5},
Jerod~Parrent\altaffilmark{2}, 
Jay Strader\altaffilmark{1},
Carles~Badenes\altaffilmark{6}
Claes~Fransson\altaffilmark{7}
Atish Kamble\altaffilmark{2},
Raffaella~Margutti\altaffilmark{8},
Michael~P.~Rupen\altaffilmark{9}, \&
Joshua~D.~Simon\altaffilmark{10}}
\altaffiltext{1}{Department of Physics and Astronomy, Michigan State University, East Lansing, MI 48824, USA}
\altaffiltext{2}{Harvard-Smithsonian Center for Astrophysics, 60 Garden Street, Cambridge, MA 02138, USA}
\altaffiltext{3}{Department of Astronomy, University of Virginia, PO Box 400325, Charlottesville, VA 22904, USA}
\altaffiltext{4}{Astronomy Department, University of Illinois at Urbana-Champaign, 1002 W.\ Green Street, Urbana, IL 61801, USA}
\altaffiltext{5}{Department of Physics, University of Illinois at Urbana-Champaign, 1110 W.\ Green Street, Urbana, IL 61801, USA}
\altaffiltext{6}{Department of Physics and Astronomy \& Pittsburgh Particle Physics, Astrophysics, and Cosmology Center (PITT-PACC), University of Pittsburgh, Pittsburgh, PA 15260, USA}
\altaffiltext{7}{Department of Astronomy, Stockholm University, AlbaNova, SE-106 91 Stockholm, Sweden}
\altaffiltext{8}{Center for Cosmology and Particle Physics, New York University, 4 Washington Place, New York, NY 10003, USA}
\altaffiltext{9}{National Research Council of Canada, Herzberg Astronomy and Astrophysics Programs, Dominion Radio Astrophysical Observatory, P.O. Box 248, Penticton, BC V2A 6J9, Canada}
\altaffiltext{10}{Carnegie Observatories, 813 Santa Barbara St., Pasadena, CA 91101, USA}
\altaffiltext{11}{chomiuk@pa.msu.edu}

\begin{abstract}
Searches for circumstellar material around Type Ia supernovae (SNe Ia) are one of the most powerful tests of the nature of SN Ia progenitors, and radio observations provide a particularly sensitive probe of this material. Here we report radio observations for SNe~Ia and their lower-luminosity thermonuclear cousins. We present the largest, most sensitive, and spectroscopically diverse study of prompt ($\Delta t \lesssim 1$ yr) radio observations of 85 thermonuclear SNe, including 25 obtained by our team with the unprecedented depth of the Karl G.\ Jansky Very Large Array. With these observations, SN\,2012cg joins SN\,2011fe and SN\,2014J as a SN~Ia with remarkably deep radio limits and excellent temporal coverage (six epochs, spanning 5--216 days after explosion, yielding $\dot{M}/v_w \lesssim 5 \times 10^{-9}~{{{\rm M}_{\odot}\ {\rm yr}^{-1}} \over {{\rm 100\ km\ s}^{-1}}}$, assuming $\epsilon_B = 0.1$ and $\epsilon_e = 0.1$). 

All observations yield non-detections, placing strong constraints on the presence of circumstellar material. We present analytical models for the temporal and spectral evolution of prompt radio emission from thermonuclear SNe as expected from interaction with either wind-stratified or uniform density media.  These models allow us to constrain the progenitor mass loss rates, with limits ranging from $\dot{M}\lesssim 10^{-9}-10^{-4}$ M$_{\odot}$~yr$^{-1}$, assuming a wind velocity $v_w=100$ km s$^{-1}$.  We compare our radio constraints with measurements of Galactic symbiotic binaries to conclude that $\lesssim$10\% of thermonuclear SNe have red giant companions. 
\end{abstract}


\section{Introduction}
\label{sec:intro}
The nature of the progenitors of Type Ia supernovae (SNe~Ia) and other thermonuclear SNe (SNe Iax, Ca-rich SNe; SN\,2002es-like explosions; \citealt{Perets_etal10, Ganeshalingam_etal12, Foley_etal13}) remains an important gap in our understanding of astrophysics, with important implications for cosmology, stellar phenomenology, and chemical evolution. There is general agreement that thermonuclear SNe involve a white dwarf in a binary system, but it remains unknown if the white dwarf is destabilized by reaching the Chandrasekhar mass, or if significantly sub-Chandrasekhar white dwarfs can explode as thermonuclear SNe. In addition, the nature of the binary companion remains a puzzle: it could be a red (super)giant, a subgiant, a main sequence star, or another white dwarf \citep{Iben_Tutukov84, Webbink84}. A wealth of review articles summarize these outstanding mysteries; see \cite{Branch_etal95, Livio01, Howell11, Wang_Han12, Maoz_etal13}. 

\subsection{Techniques for testing progenitor scenarios}
Unlike H-rich core-collapse supernovae, which have luminous massive star progenitors \citep{Smartt09, Smartt15}, the direct detection of the themonuclear SN progenitor system in pre-explosion images is not usually a viable technique. The expected white dwarf and any likely companion are simply too faint for pre-explosion measurements to be constraining \citep[e.g.,][]{Maoz_Mannucci08}. Noteable exceptions are the two nearest SNe~Ia of recent times, SN\,2011fe and SN\,2014J (these also yielded non-detections; \citealt{Li_etal11b, Kelly_etal14}) and SNe~Iax (see below).

Therefore, indirect techniques are needed for constraining the progenitors of thermonuclear SNe. Such tests include:\\
{\bf (a)} the impact of the companion on the early SN light curve \citep{Kasen10, Hayden_etal10a, Bianco_etal11, Ganeshalingam_etal11, Brown_etal12, Foley_etal12b, Zheng_etal13, Olling_etal15, Marion_etal15}; \\
{\bf (b)} searches for the companion star in nearby SN~Ia remnants \citep{Ruiz-Lapuente_etal04, Kerzendorf_etal09, Kerzendorf_etal12, Kerzendorf_etal13, Kerzendorf_etal14, Schaefer_Pagnotta12}; \\
{\bf (c)} late-time nebular spectroscopy to search for companion material entrained in the SN ejecta \citep{Leonard07, Shappee_etal13, Lundqvist_etal13, Lundqvist_etal15};\\
{\bf (d)} estimates of white dwarf mass and ejecta mass from light curves and nucleosynthesis \citep{Howell_etal06, Stritzinger_etal06, Seitenzahl_etal13, Scalzo_etal14, Yamaguchi_etal15}; and\\
{\bf (e)} characterization of the circumstellar environment at the SN site  \citet{Panagia_etal06, Russell_Immler12, Margutti_etal12, Chomiuk_etal12, Margutti_etal14}. The latter strategy, carried out using radio continuum emission, is the subject of this paper.

\subsection{Studies of circumstellar material}

\subsubsection{Theoretical expectations}
The various progenitor channels associated with thermonuclear SNe are each associated with a signature circumbinary environment (see \citealt{Chomiuk_etal12} and \citealt{Margutti_etal14} for more discussion).  The composition, density, and radial profile are shaped by the late time evolution of the progenitor system. For example, the densest circumstellar medium (CSM) should be expelled by a red giant companion; the SN would then explode in a wind-stratified medium, with the density declining with radius from the giant. White dwarfs with main-sequence or sub-giant companions would be surrounded by a much less dense wind, but may host dense narrow shells of material at some radii: the remnants of nova explosions \citep{Chomiuk_etal12}. Theory predicts a diversity of circumbinary environments for double degenerate systems (where the companion is a He or CO white dwarf), including wind-stratified material \citep{Guillochon_etal10, Dan_etal11, Ji_etal13}, nova shells \citep{Shen_etal13}, tidal tails of shredded white dwarf material \citep{Raskin_Kasen13}, or low densities indistinguishable from the local interstellar medium. Therefore, the overarching goal of the radio survey presented here is the search for material in the local explosion environment, with important implications for binary evolution and mass transfer in thermonuclear SN progenitor systems.
                                   
A wide range of multi-wavelength observations have, to date, placed important constraints on the environments of SNe~Ia. Authors typically constrain the circumbinary environment assuming a wind profile centered at the SN site. Density then scales as the progenitor mass loss rate ($\dot{M}$) and the inverse of the wind velocity ($v_w$) as $\rho_{\rm CSM}=\dot{M}/(4\pi r^2 v_w)$; both $\dot{M}$ and $v_w$ are assumed to be constant in the years leading up to explosion. Spherical symmetry of both the explosion and CSM are typically assumed.

\subsubsection{Prompt H$\alpha$ observations}
One of the most straight-forward methods for detecting CSM is the search for optical emission lines from the wind that has been ionized and heated by the shock radiation. Limits from relatively early H$\alpha$ non-detections for several individual ``normal" SNe~Ia imply $\dot{M}/v_w\lesssim 10^{-4}~{{{\rm M}_{\odot}\ {\rm yr}^{-1}} \over {{\rm 100\ km\ s}^{-1}}}$ \citep{Cumming_etal96, Mattila_etal05} (later-time nebular constraints on H$\alpha$ provide a parallel test of thermonuclear SN progenitors, constraining the amount of H-rich material entrained in the SN ejecta, presumably from the blast wave sweeping over a companion star; \citealt{Leonard07, Shappee_etal13, Lundqvist_etal13, Lundqvist_etal15}).

\subsubsection{Radio observations}
Observations of non-thermal radio emission trace electrons accelerated to relativistic speeds via diffusive shock acceleration, emitting synchrotron radiation. Radio observations with the Very Large Array over a period of two decades set upper limits on the emission from 27 SNe~Ia (\citealt{Panagia_etal06}, and references therein). The most constraining luminosity limits are $\lesssim 10^{25}$ erg s$^{-1}$ Hz$^{-1}$ over timescales of 10--1000 days after explosion. In order to predict the expected radio luminosity, \citet{Panagia_etal06} assume a model in which the radio-emitting SN blast wave is expanding at 10,000 km s$^{-1}$, the synchrotron emission is absorbed by free-free processes in an external wind of temperature of $2 \times 10^{4}$ K, and the synchrotron luminosity is scaled to observations of the Type Ib/c supernova SN\,1983N. Typical limits on the mass loss rate are then $\dot{M}/v_w\lesssim 10^{-5}~{{{\rm M}_{\odot}\ {\rm yr}^{-1}} \over {{\rm 100\ km\ s}^{-1}}}$, but extend down to $\dot{M}/v_w\lesssim 3\times 10^{-7}~{{{\rm M}_{\odot}\ {\rm yr}^{-1}} \over {{\rm 100\ km\ s}^{-1}}}$ in the most sensitive cases. 

The radio observations from \citet{Panagia_etal06} were stacked by \citet{Hancock_etal11} to achieve maximal sensitivity. The stacked radio observation still yielded a non-detection, but the ``typical" limit on the progenitor mass loss rate is an order of magnitude more constraining, $\dot{M}/v_w\lesssim 10^{-6}~{{{\rm M}_{\odot}\ {\rm yr}^{-1}} \over {{\rm 100\ km\ s}^{-1}}}$ (using the formalism of \citealt{Panagia_etal06}, which assume a fixed blast wave velocity and free-free absorption). 

More recently, very deep radio limits have been placed on the nearby SNe~Ia 2011fe and 2014J using the synchrotron self-absorption formalism described in detail in this paper \citep{Horesh_etal11, Chomiuk_etal12, Perez-Torres_etal14}. This reconsideration of synchrotron opacity was driven by the work of \citet{Chevalier98}, which showed that for the fast blast wave velocities and relatively low-density environments of Type~I SNe, synchrotron self-absorption dominates over free-free absorption. Due to the proximity of SN\,2011fe and SN\,2014J and intensive observing campaigns, mass loss rates in these two explosions are strongly constrained, $\dot{M}/v_w\lesssim 7\times 10^{-10}~{{{\rm M}_{\odot}\ {\rm yr}^{-1}} \over {{\rm 100\ km\ s}^{-1}}}$.

\subsubsection{X-ray observations}
Upper limits on the X-ray emission from thermonuclear SNe also constrain the density of CSM, although two different X-ray emission mechanisms have been assumed in the literature, affecting the translation from X-ray luminosity to circumstellar density. While  \citet{Hughes_etal07} and \citet{Russell_Immler12} assume the X-ray emission is thermal bremsstrahlung from the hot SN-shocked material, \citet{Chevalier_Fransson06} argue that Inverse Compton radiation should dominate the X-ray luminosity in Type I supernovae, produced when shock-accelerated relativistic electrons interact with the SN photospheric emission. 

X-ray limits presented for 53 thermonuclear SNe by \citet{Russell_Immler12} imply $\dot{M}/v_w\lesssim 10^{-5}~{{{\rm M}_{\odot}\ {\rm yr}^{-1}} \over {{\rm 100\ km\ s}^{-1}}}$, assuming thermal radiation. However, Inverse Compton radiation is expected to be significantly brighter than thermal emission in thermonuclear SNe, at the expected low circumstellar densities and within the first $\sim$40 days following explosion, so the limits presented by \citet{Russell_Immler12} would be more constraining if an Inverse Compton model was assumed. \citet{Margutti_etal12, Margutti_etal14} obtained deep X-ray observations of the nearby SNe~Ia 2011fe and 2014J and obtained limits on the mass loss rates comparable to those from the radio, $\dot{M}/v_w\lesssim 10^{-9}~{{{\rm M}_{\odot}\ {\rm yr}^{-1}} \over {{\rm 100\ km\ s}^{-1}}}$.

\subsubsection{Late-time observations of SN remnants}
Radio and X-ray signatures probe the circumstellar material at the present location of the blast wave, so observations in the first year (as presented in the current study) trace the properties of the CSM at radii $\sim10^{14}-10^{17}$ cm. Observations of SN remnants (observed hundreds to thousands of years after explosion) are complementary, constraining the CSM on $\sim$pc scales. Studies of SN remnants show that their dynamical properties are compatible with an interaction with a uniform ambient medium, with densities typical of the warm phase of the ISM \citep{Badenes_etal07, Yamaguchi_etal14}. These observations rule out large energy-driven cavities around SNe Ia (with the single exception being SN remnant RCW 86, which appears to be expanding into a cavity; \citealt{Badenes_etal07, Williams_etal11, Broersen_etal14, Yamaguchi_etal14}). Such cavities are predicted products of fast outflows from the progenitor system ($>$few $\times 100$ km s$^{-1}$), because above a certain critical outflow velocity, radiative losses do not affect the shocked material, and the wind will excavate a low-density energy-driven cavity \citep{Koo_McKee92a, Koo_Mckee92b}. Therefore, the large-scale structure of the CSM around a SN Ia progenitor will depend on the properties of the outflow, specifically whether it was faster or slower than the critical velocity, and whether it was active at the time of the explosion (see \citealt{Badenes_etal07}). 

\subsubsection{\ion{Na}{1}~D absorption features}
Another observational tracer of CSM---\ion{Na}{1}~D optical absorption features observed against the SN continuum---probes the presence of smaller structures (e.g., shells, clumps) around thermonuclear SNe. \ion{Na}{1}~D absorption is sensitive to CSM at a range of distances from the SN site, as long as the material is positioned along our line of sight. This tracer is returning more than the simple non-detections repeatedly observed at radio, X-ray, and H$\alpha$ wavelengths. \ion{Na}{1}~D absorption profiles that temporally vary after explosion indicate material local to the SN site ($\sim10^{16}-10^{18}$ cm away), and such variations have been observed in a handful of normal SNe~Ia \citep{Patat_etal07a, Blondin_etal09, Simon_etal09}. In addition, statistical studies using single-epoch \ion{Na}{1}~D profiles for larger samples of SNe~Ia find that blue-shifted absorption is over-represented in SNe~Ia, implying that SNe~Ia host outflowing CSM along our line of sight in approximately a quarter of cases \citep{Sternberg_etal11, Foley_etal12, Maguire_etal13, Phillips_etal13}. 

In the few SNe where time-variable \ion{Na}{1}~D is observed, the absorbing material appears consistent with nova shells which have plowed into and swept up a red giant wind \citep{Wood-Vasey_Sokoloski06, Patat_etal11a}, or an old nova shell expanding into the interstellar medium and decelerating \citep{Moore_Bildsten12, Shen_etal13}. In the case of SN\,2006X, the first SN~Ia to show time-variable absorption, deep radio limits were observed with the pre-upgrade VLA in the months following explosion, which should further constrain the radial distribution of material around the SN \citep{Patat_etal07a, Chugai08}. However, to date, it remains unclear if the radio non-detections in SN\,2006X are consistent with the locations of shells inferred from the \ion{Na}{1}~D absorption.

\subsection{A diversity of progenitors?}
The detection of time-variable or blue-shifted \ion{Na}{1}~D absorption around some, but not all, SNe~Ia, and an observed correlation with explosion properties \citep{Foley_etal12}, has led to widespread speculation that there is likely more than one progenitor channel to SN~Ia explosions. Further stoking this speculation, there is now an established class of Type Ia-CSM SNe, which show strong H$\alpha$ emission implying dense surroundings \citep{Hamuy_etal03, Dilday_etal12, Silverman_etal13}.  The circumstellar interaction in these systems is so strong that it contributes to the optical luminosity itself; continuum light curve modeling implies a very high mass loss rate of $\dot{M}/v_w \approx 10^{-1}~{{{\rm M}_{\odot}\ {\rm yr}^{-1}} \over {{\rm 100\ km\ s}^{-1}}}$. SNe~Ia-CSM are rare and have therefore only been observed at large distances ($\gtrsim 100$ Mpc).

In addition, there have been some recent successes at identifying progenitors of low-luminosity thermonuclear SNe (which we consider here in three distinct sub-types; Ca-rich SNe, SNe Iax, and SN\,2002cs-like explosions; \citealt{Perets_etal10, Ganeshalingam_etal12, Foley_etal13, White_etal15}). These sub-types may each have a distinct progenitor channel; for example, \citet{Foley_etal13} claim that SNe~Iax marking the deflagration of a CO white dwarf after it accretes from a He star companion. This claim is further supported by a detection of the progenitor system of SN\,2012Z in pre-explosion {\it HST} images (\citealt{McCully_etal14}. The blue color and luminosity of the progenitor system are consistent with the Galactic binary V445~Pup, which hosted a He nova and is likely composed of a He star and white dwarf (e.g., \citealt{Ashok_Banerjee09,Goranskij_etal10}). There is also a possible post-explosion detection of the  companion or bound remnant of SN~Iax 2008ha \citep{Foley_etal14b}; the red color of this detection is strikingly different than observed in SN\,2012Z, implying that even within the sub-type of SNe~Iax, the progenitor systems may be diverse (see also \citealt{Foley_etal15}). Further evidence that at least some lower-luminosity thermonuclear SNe have non-degenerate companions comes from an early-time blue excess in the ultraviolet/optical light curve of the SN\,2002es-like iPTF14atg, consistent with models of SN ejecta interacting with a companion star \citep{Cao_etal15}.

\subsection{This paper}
Today, radio observations remain one of the most sensitive and efficient tracers of the environments of thermonuclear SNe. While X-ray observations can obtain similar sensitivities in some cases, X-ray observations are much more expensive than radio observations, as they require longer exposure times. While radio observations require a fair amount of modeling to interpret, such models can be tested with well-studied radio bright Type Ib/c SNe \citep{Chevalier_Fransson06}. Once a model formalism is developed, radio observations with upgraded facilities place strong constraints on CSM around nearby thermonuclear SNe. In addition, and in contrast to \ion{Na}{1}~D measurements, radio observations constrain the CSM at a well-known radius, enabling studies of the radial profile of material around thermonuclear SNe. 

In \S 2, we discuss our sample of 85 thermonuclear SNe and divide them into nine spectroscopically-determined sub-types. In \S 3, we present new radio observations of 25 nearby thermonuclear SNe, observed within the first year following explosion using the expanded bandwidth and factor of $\sim$5 improvement in sensitivity of the Karl G.\ Jansky Very Large Array (VLA).  We combine them with published and archival radio observations obtained during the first year after explosion. We then describe our self-consistent model for the synchrotron emission from exploding white dwarfs in \S 4. In \S 5, we apply this model to our dataset to constrain the density of the circumstellar environments on scales $r\lesssim 10^{17}$ cm. We discuss specific sources of note and conclusions for individual sub-types in \S 6, and we conclude in \S 7.


\section{Supernova sample}
\label{sec:sample}

\subsection{Sample selection}
\begin{figure}
\centerline{\includegraphics[width=2.9in,angle=90]{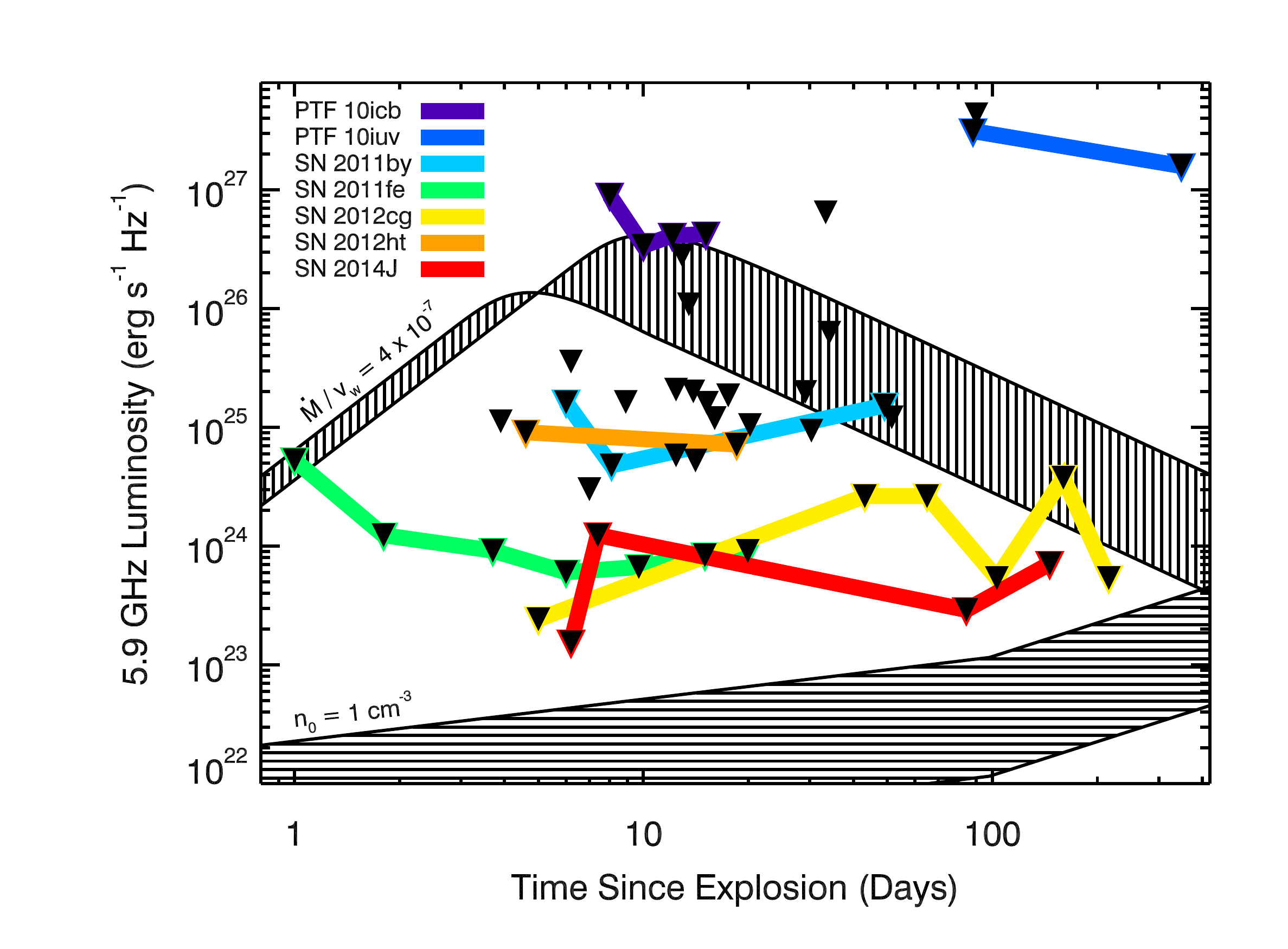}}
\vspace{-0.1in}
\caption{5.9 GHz upper limits from the Jansky VLA (downward facing black triangles), compared with models for a thermonuclear SN interacting with CSM (black hatched regions). SNe that are observed at multiple epochs are connected with colored solid lines (see Tables \ref{snobs} and \ref{snobs_pan}). Observations that were carried out at frequencies differing from 5.9 GHz were normalized to 5.9 GHz assuming an optically-thin synchrotron spectral index, $\alpha = -1$. The model light curve that is vertically hatched and peaks around Day 10 is for a wind-stratified CSM with $\dot{M}/v_w = 4\times 10^{-7}~{{{\rm M}_{\odot}\ {\rm yr}^{-1}} \over {{\rm 100\ km\ s}^{-1}}}$ assuming $\epsilon_e = 0.1$, $M_{\rm ej,Ch}=1$, and $E_{\rm K,51}=1$; its finite width shows the effect of varying $\epsilon_B$ between 0.01--0.1. The horizontally hatched light curve model that monotonically increases is for a uniform density CSM with $n_0 = 1$ cm$^{-3}$, $\epsilon_e = 0.1$, $M_{\rm ej,Ch}=1$, and $E_{\rm K,51}=1$---again, its width is set to span $\epsilon_B = 0.01-0.1$.}
\label{fig:evla}
\vspace{0.1in}
\end{figure}

We observed 25 thermonuclear SNe with the Karl G.\ Jansky VLA during its first five years of operation, with a goal of observing as soon as possible after SN spectroscopic identification to obtain the most sensitive limits on wind-stratified CSM (Figure \ref{fig:evla}). These VLA-observed SNe were selected to be nearby ($d\lesssim 30$ Mpc), discovered before optical maximum, and accessible to the VLA (declination $\gtrsim -35^{\circ}$; Table \ref{snpar}). One clear exception to these criteria is PTF\,10guz at a distance of $\sim$740 Mpc, which was observed as part of a program following up unusual transients (program ID AS1020), as PTF\,10guz was deemed a likely super-Chandrasekhar SN~Ia \citep{Nugent_etal10_10guz}. Here we present radio observations acquired in the first year following explosion (Table \ref{snobs}); later-time VLA observations for many thermonuclear SNe will be presented in a forthcoming paper (Chomiuk et al.\ 2016, in preparation). 

We further present archival observations of a sample of 31 older, mostly unpublished thermonuclear SNe observed with the historical VLA, obtained between 1998--2009 (Table \ref{snobs_old}). The VLA archive was searched for observations of nearby, bright, or particularly interesting thermonuclear SNe obtained within one year of explosion and providing more than 5 minutes on source. In particular, we searched the VLA archive for observations of (i) all SNe~Ia listed on the Rochester Astronomy Supernova page as discovered between 1995--2009 and $\lesssim$15 mag at optical maximum; (ii) all SNe~Iax catalogued by \citet{Foley_etal13}; (iii) all Ca-rich SNe identified in \citet{Perets_etal10}; (iv) SNe~Ia potentially exceeding the Chandrasekhar mass as listed in \citet{Taubenberger_etal11}; and (v) ``non-PTF" SNe~Ia-CSM as catalogued by \citet{Silverman_etal13}.

Finally, we re-consider 29 SNe~Ia with radio limits observed with the VLA in the first year following explosion, published in the refereed literature (Table \ref{snobs_pan}). 26 of these are from \citet{Panagia_etal06}, which tend to be bright ($\leq$ 14 mag at maximum) events that occurred over the period 1980--2003. Limits on three additional events of particular interest were published in \citet{Kasliwal_etal12, Dilday_etal12}, and \citet{Cao_etal15}; these limits are also plotted in Figure \ref{fig:evla}, as they were obtained with the upgraded Jansky VLA.

The complete sample of 85 thermonuclear SNe observed with the VLA is presented in Table \ref{snpar}. The SN positions listed in Table \ref{snpar} originate from the ``List of Supernovae" published by the IAU Central Bureau for Astronomical Telegrams. Host galaxy distances and morphological types are from the NASA Extragalactic Database (NED). We used the average ``redshift-independent distance'' measurement in NED if available as of June 2015; otherwise, we used the Galactocentric GSR distance from NED, which assumes $H_{0} = 73$ km s$^{-1}$ Mpc$^{-1}$. For three particularly nearby SNe~Ia, high-quality distance estimates originating from specific publications were used---SN\,2011fe in M\,101 \citep{Shappee_Stanek11}, SN\,2014J in M\,82 \citep{Dalcanton_etal09}, and SN\,1986G in NGC\,5128 \citep{Harris_etal10}. The dates of optical $B$-band maximum for SNe were estimated from published data, as referenced in Table \ref{snpar}. Where public light curves are not available, we estimate the time of explosion from spectral information (occasionally from classification circulars).

\subsection{Categorization of supernovae}
\label{sec:cat}
The spectroscopic diversity of thermonuclear SNe is well established and extensively studied  \citep[e.g.,][]{Branch_etal06, Perets_etal10, Foley_etal13, Parrent_etal14, White_etal15}. Recently, there has arisen evidence for a diversity of progenitors, and the nature of SN progenitors is likely aligned with spectroscopic categorization \citep[e.g.,][]{Perets_etal10, Foley_etal12, Foley_etal13, Maguire_etal13}. Therefore, we divide the thermonuclear SNe observed with the VLA into nine categories, to provide limits on the radio emission from particular sub-types:
\begin{enumerate}[i.]
\item``Iax" low-luminosity explosions that resemble SN\,2002cx \citep{Foley_etal13};
\item Low-luminosity events of which SN\,2002es is the prototype \citep{Ganeshalingam_etal12};
\item``Ca-rich" low-luminosity events of which SN\,2005E is the prototype \citep{Perets_etal10, Kasliwal_etal12}; 
\item``Cool"/faint SNe~Ia that resemble SN\,1991bg \citep{Benetti_etal05, Branch_etal06}; 
\item``Broad-line"/high velocity SNe~Ia, which display a high temporal velocity gradient as defined by \citet{Benetti_etal05}. SN\,1984A is a classic example \citep{Branch_etal06};
\item``Core-normal" events with a low temporal velocity gradient, of which SN\,1994D is the prototype \citep{Benetti_etal05, Branch_etal06}; 
\item ``Shallow-Silicon" SN\,1999aa-like SNe~Ia, which also show low temporal velocity gradients \citep{Benetti_etal05, Branch_etal06};
\item Luminous SNe~Ia which are candidate super-Chandrasekhar-mass explosions (e.g., SN\,2003fg; \citealt{Howell_etal06}); 
\item ``Ia-CSM", which are SNe~Ia that develop hydrogen emission lines, signalling circumstellar interaction (e.g., SN\,2005gj; \citealt{Silverman_etal13}).  
\end{enumerate}
We assigned each VLA-observed thermonuclear SN one of these categories, based upon published spectra and circulars. The thermonuclear SNe are listed under their respective categories in Table \ref{snpar}.  

In some cases (e.g., when spectroscopy was only obtained after maximum light), it was difficult to differentiate between core-normal and broad-line classifications. In these cases, we categorized the SNe as core-normal and marked them with a superscript in Table \ref{snpar}. In addition, several of the SNe in our sample are hybrid cool/normal SNe~Ia; these are also distinguished in Table \ref{snpar} with a superscript. 

While the time of optical maximum light is usually well measured, it is the time since explosion that defines the properties of the radio evolution.  To this end, we estimate the explosion time with respect to maximum light (i.e. the rise time of the $B-$band light curve) separately for the different classes as:
\begin{itemize}
\item Iax: $t_{\rm rise} \approx$15 days \citep{Foley_etal13, Stritzinger_etal14};
\item 02es-like: we only consider one member of this class---iPTF\,14atg---and its rise time was measured to be 18.8 days \citep{Cao_etal15};
\item Ca-rich: $t_{\rm rise} \approx $12 days \citep{Kasliwal_etal12};
\item Cool: $t_{\rm rise} \approx 13$ days \citep{Ganeshalingam_etal11};
\item Broad-line: $t_{\rm rise} \approx 16.6$ days \citep{Ganeshalingam_etal11}; 
\item Core-normal: $t_{\rm rise} \approx 18.0$ days \citep{Ganeshalingam_etal11};
\item Shallow-silicon: $t_{\rm rise} \approx 18.1$ days \citep{Ganeshalingam_etal11}; 
\item Super-Chandrasekhar: $t_{\rm rise} \approx 24$ days \citep{Scalzo_etal10, Silverman_etal11};
\item Ia-CSM: $t_{\rm rise} \approx 30$ days \citep{Silverman_etal13}.
\end{itemize}
To calculate the time elapsed between explosion and VLA observation (as listed in Tables \ref{snobs}--\ref{snobs_pan}), we utilize the estimates of $B$-band optical maximum from Table 1, and these type-dependent rise times.

\section{Observations}
\label{sec:obs}

Observations with the upgraded VLA (2010 and after) were obtained in the standard continuum observing mode. During the early VLA commissioning phase, this provided a bandwidth of $2\times128$ MHz, and later $2\times1024$ MHz of bandwidth (programs AS1015, AS1058, 11B-177, 11B-217, 12A-482, 13B-454, and TOBS0008).  The details of the observations are presented in Table \ref{snobs}. Most observations were obtained at C band (4--8 GHz). We utilized the maximum bandwidth available at the time of each observation (see Table \ref{snobs}) and recorded four polarization products.  The observations were typically 1 hr in total duration resulting in $\sim 35-40$ min on source. For phase calibration, we used a complex gain calibrator source within $\sim 10$ degrees of each SN. For flux and bandpass calibration, one of 3C48, 3C147, or 3C286 was observed in each block. 
 
 The data were edited, calibrated, and imaged using standard routines in the Astronomical Image Processing System (AIPS; \citealt{Greisen03}).  Each sideband was imaged individually. Individual sidebands were checked for a detection at the SN position, and when none was found, maximal sensitivity was achieved by smoothing the higher-frequency image to the resolution of the lower-frequency image, and then averaging the images from the two sidebands together.  For those SNe observed in more compact configurations and located near regions of detectable diffuse host galaxy emission, we used a limited {\it uv} range in imaging the SN field to reduce contamination on short baselines. In some cases where there was significant contaminating flux from a nearby source, we chose to only use the higher-frequency sideband; these are marked in Table \ref{snobs} as observations with diminished bandwidth (i.e., PTF\,10icb, SN\,2011iv, SN\,2011iy, SN\,2014J). Our reduced images typically reach expectations for the theoretical noise limit, except in cases where there are bright continuum sources nearby.

Observations with the pre-upgrade (pre-2010) VLA span frequencies 1.4--22.5 GHz, but were mostly commonly obtained at 4.8 or 8.4 GHz to maximize sensitivity (Tables \ref{snobs_old} and \ref{snobs_pan}). They were acquired in standard continuum mode ($2\times43$ MHz bandwidth). These observations are typically $\sim$5 times less sensitive than the post-upgrade observations, and often tackle thermonuclear SNe at greater distances, so they are significantly less constraining than the post-upgrade VLA observations in terms of radio luminosity. 

We did not detect positionally coincident radio emission from any of the thermonuclear SNe in our sample.  To determine upper limits on the radio flux density, we measured the image rms noise ($\sigma_{\nu}$) and the flux density ($S_{\nu}$) at the optical SN position. Upper limits considered in this study are of 3$\sigma$ significance. In cases where the flux density measured at the SN position is a positive value, we calculate the flux density upper limit as $S_{\nu} + 3\sigma_{\nu}$; where the flux density at the SN position is negative, we set the flux density upper limit to $3\sigma_{\nu}$. Limits taken from the literature, as listed it Table \ref{snobs_pan}, are simply $3\sigma_{\nu}$, as these publications do not provide flux densities at the positions of the SNe. Luminosity upper limits presented in Tables \ref{snobs}, \ref{snobs_old}, and \ref{snobs_pan} assume the distances listed in Table \ref{snpar}. These luminosity constraints are translated into upper limits on the density of CSM (also listed in Tables \ref{snobs}--\ref{snobs_pan}) as described in Section \ref{sec:modcomp}.

\section{Synchrotron emission from thermonuclear supernovae}
\label{sec:model}

While optical observations of young freely-expanding SNe probe the expanding ejecta, radio observations trace the dynamical interaction of the fastest ejecta with the circumstellar environment.  This interaction accelerates particles to relativistic speeds and amplifies the magnetic field in the shocked region, producing radio synchrotron emission \citep{Chevalier82b}. We make the common assumption that the energy densities in both relativistic particles and magnetic field scale with the post-shock energy density,  $\sim \rho_{\rm CSM}\, v_s^2$, where $\rho_{\rm CSM}$ is the mass density of the CSM being impacted by the shock at time of observation, and $v_s$ is the blast wave velocity \citep{Chevalier96}. Therefore, a SN will be more radio luminous if it has a fast shock velocity or expands into CSM of high density.

This interaction between SN blast and the surrounding medium leads to a double shock system in which the forward shock plows through the CSM while the reverse shock travels into and decelerates the ejecta \citep{Chevalier82b}.  A contact discontinuity at radius, $R_c$, divides the forward-shocked CSM from the reverse-shocked ejecta, and the total shocked region at any time is bounded by the locations of the forward and reverse shocks.  The temporal evolution of $R_c$ is described by a self-similar solution determined by the density profile of the SN ejecta, $\rho_{\rm SN}$, and that of the local CSM, $\rho_{\rm CSM}$ \citep{Chevalier82a}.

The radio properties of thermonuclear SNe may be reasonably predicted from radio observations and detailed modeling of well-observed SNe~Ib/c. These explosions are produced through the gravitational collapse of massive stars that have ejected their hydrogen envelopes prior to explosion \citep{Chevalier_Fransson06}.  Since both white dwarfs and Type Ib/c progenitors lack a massive stellar envelope, the blast waves of both thermonuclear SNe and Ib/c are unencumbered and reach similar high velocities of $\overline{v}_s\approx 0.1c-0.3c$. As described below, making the reasonable assumption that the properties of accelerated electrons and amplified magnetic field are similar for thermonuclear SNe and Ib/c explosions, we construct models for the radio emission from SNe~Ia and other white dwarf SNe. 

\subsection{A model for the ejecta profile}

\begin{figure}
\centerline{\includegraphics[width=2.8in,angle=270]{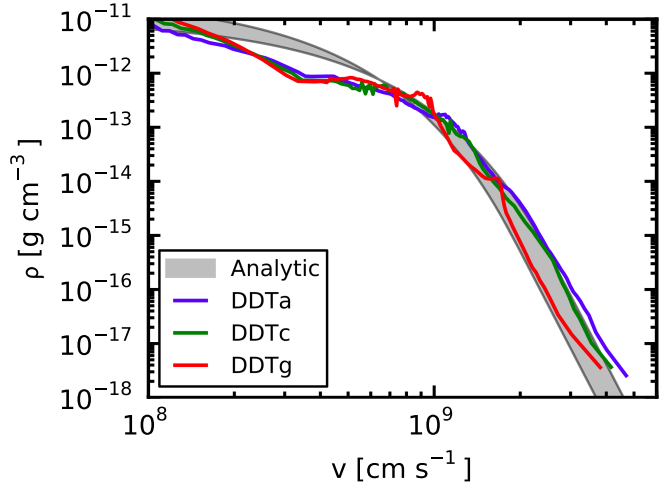}}
\caption{Our analytic form of the radial density profile for SN~Ia ejecta (grey band), compared with ejecta profiles from SN~Ia simulations (colored lines). The width of the grey band spans $E_{\rm K,51} = 0.8-1.5$ and assumes $M_{\rm ej,Ch}=1$. The red, green, and blue lines represent one-dimensional hydrodynamic+nucleosythesis simulations of SNe~Ia from \citet{Badenes_etal03, Badenes_etal05, Badenes_etal08}. They correspond to delayed detonations with energies, $E_{\rm K,51} = 0.8$ (DDTg; red), $E_{\rm K,51} = 1.2$ (DDTc; green), and $E_{\rm K,51} = 1.4$ (DDTa; blue); all simulations assume $M_{\rm ej,Ch}=1$. These profiles all correspond to $10^6$~s after explosion, assuming free expansion and no CSM interaction.}
\label{fig:ejecta}
\vspace{0.1in}
\end{figure}

The unshocked SN ejecta expand freely and homologously, such that the expansion velocity ($v$) at some radius ($r$) in the ejecta observed at a time ($t$) after explosion is $v = r/t$. The early photospheric emission from SNe~Ia points to an exponential density distribution for the inner, denser ejecta layers of exploded white dwarfs, $\rho_{\rm SN} ({\rm inner})\propto {\rm e}^{-v}$ (i.e., the W7 model; \citealt{Nomoto_etal84}).  This profile is supported by observations of the reverse shock-heated ejecta in Type I Galactic SN remnants (e.g., \citealt{Hamilton_Fesen88, Itoh_etal88, Dwarkadas_Chevalier98}). Meanwhile, a steep power-law density profile has been suggested for the outer ejecta layers, $\rho_{\rm SN}({\rm outer}) \propto v^{-n}$, based on theoretical expectations for the explosion of a polytrope \citep{Colgate_McKee69, Chevalier82b, Matzner_McKee99, Nakar_Sari10}.  Combining these two profiles, we construct a radial density profile for thermonuclear SN ejecta characterized by a relatively flat, exponential density profile at low velocities (inner ejecta) and a steep power-law density profile at high velocities (outer ejecta).  A break velocity, $v_b$, divides the two regimes. Our analytic model for the ejecta density profile matches the results of SN~Ia simulations well (Figure \ref{fig:ejecta}). 

We adopt the formalism of \citet{Dwarkadas_Chevalier98} for the inner ejecta profile,
\begin{equation}
\rho_{\rm SN}({\rm inner})=\rho_{0,{\rm in}}\ e^{-v/{v_e}}\ t^{-3}
\end{equation}
where 
\begin{equation}
\rho_{0,{\rm in}}\approx \left(7.67\times 10^6~\rm g~cm^{-3}\right) M_{\rm ej,Ch}^{5/2}~E_{\rm K,51}^{-3/2}
\end{equation}
 \begin{equation}
v_e\approx \left(2.44\times 10^8~\rm cm~s^{-1}\right)~M_{\rm ej,Ch}^{-1/2}~E_{\rm K,51}^{1/2}.
\end{equation}
Here, $E_{\rm K,51}$ is the kinetic energy of the SN in units of $10^{51}$ erg, and $M_{\rm ej,Ch}$ is the ejecta mass in units of the Chandrasekhar mass, 1.4 M$_{\odot}$.

We characterize the outer layers of the ejecta with a power-law profile appropriate for a compact progenitor star derived from the harmonic mean models of  \citet[][Equation 46 and Table 5]{Matzner_McKee99}. The case of interest here is a $\gamma = 4/3$ polytrope, which is appropriate to a relativistic white dwarf approaching the Chandrasekhar mass.
In this framework, 
\begin{equation}
\rho_{\rm SN}({\rm outer})=\rho_{0,{\rm out}}~v^{-10.18}~t^{-3}
\label{eqn:outer}
\end{equation}
where 
\begin{equation}
\rho_{0,{\rm out}}\propto M_{\rm ej}^{-2.59}~E_{\rm K,51}^{3.59}
\end{equation}
as shown by \citet{Berger_etal02} and \citet{Chevalier_Fransson06} for Wolf-Rayet star explosions.

Equating the slopes and normalization of the profiles for the inner and outer ejecta regions  we find the location of the break velocity:
\begin{equation}
{\rm v}_b\approx (2.48\times 10^9~{\rm cm~s^{-1}} )~M_{\rm ej,Ch}^{-1/2}~E_{\rm K,51}^{1/2}
\end{equation}
and the density normalization of the outer ejecta (in cgs units):
\begin{equation}
\rho_{0,{\rm out}}\approx (1.3\times 10^{98}~{\rm g~cm^{-3}})~M_{\rm ej,Ch}^{-2.59}~E_{\rm K,51}^{3.59}.
\end{equation}
Our default assumption in this paper are the often-assumed parameters for SNe~Ia, $M_{\rm ej,Ch}=1$, $E_{\rm K,51}=1$. We also apply our model to the ejecta of sub-luminous SN~Iax and Ca-rich explosions, but note that in these cases, $E_{\rm K,51}$ and $M_{\rm ej,Ch}$ are both $<1$ \citep{Perets_etal10, Foley_etal13}; implications are discussed further in \S \ref{sec:caveat} and \S \ref{sec:indiv}. 

\subsection{Self-similar solutions for blast wave dynamics}

The initial interaction between the SN ejecta and the surrounding medium occurs at the outermost  reaches of the ejecta which expand at the highest velocities, reaching great speeds because they were accelerated by the SN shock traveling down the steep density gradient at the outer edge of the white dwarf \citep{Sakurai60}. This interaction is self-similar, as described in \citet{Chevalier82a}. We parameterize the radial density profile of the outer SN ejecta as $\rho_{\rm SN}\propto r^{-n}$, and that of the CSM as $\rho_{\rm CSM}\propto r^{-s}$.  The CSM is characterized as $\rho_{\rm CSM}=A~r^{-s}$ where $A$ is a normalization constant. Here $s=0$ in the case of a constant density medium and $s=2$ for a stellar-wind stratified environment.  

At a particular time $t$, the shock contact discontinuity is located at:
\begin{equation}
R_c=\left[\frac{C~\rho_{0,{\rm out}}}{A}\right]^{1/(n-s)} t^{(n-3)/(n-s)}
\label{eqn:rc}
\end{equation}
where $C$ is a constant that depends on the properties of the interaction region \citep{Chevalier82a}.  Self-similarity therefore imposes a strict relation between $n$, $s$ and the temporal evolution of the contact discontinuity radius, $R_c\propto t^m$, with $m=(n-3)/(n-s)$ \citep{Chevalier98}.

We consider two specific self-similar solutions associated with uniform-density and wind-stratified CSM environments, respectively.  In the case of a constant density medium, $s=0$, $C\approx 0.32$, and the ratio of the forward shock radius, $R_{FS}$, to that of the contact discontinuity is $R_{FS}/R_c\approx 1.13$ (see Table 1 of \citealt{Chevalier82a}).  For a wind-stratified CSM, $s=2$, $C\approx 0.064$, and $R_{FS}/R_c\approx 1.24$.

For $n=10.18$ (Equation \ref{eqn:outer}), we find $m\approx 0.71$ for $s=0$, and $m\approx 0.88$ for $s=2$. The radial evolution of the forward shock is thus given by
\begin{equation}
R_{FS}(s=0)\approx (14\times 10^{15}~{\rm cm})~M_{\rm ej,Ch}^{-0.25}~E_{\rm K,51}^{0.35}~n_0^{-0.10}~t_{10}^{0.71} 
\label{eqn:rs0}
\end{equation}
\noindent
\begin{equation}
R_{FS}(s=2)\approx (5.0\times 10^{15}~{\rm cm})~M_{\rm ej,Ch}^{-0.32}~E_{\rm K,51}^{0.44}~A_{\star}^{-0.12}~t_{10}^{0.88} 
\label{eqn:rs2}
\end{equation}
where $t_{10}$ is the time since explosion normalized to 10 days, $n_0$ is the CSM particle number density normalized to 1 cm$^{-3}$ in the uniform density case ($s=0$), and $A_{\star}=A/5\times 10^{11}~\rm g~cm^{-1}$. $A_{\star} = 1$ corresponds to $\dot{M}/v_w = 10^{-6}~{{{\rm M}_{\odot}\ {\rm yr}^{-1}} \over {{\rm 100\ km\ s}^{-1}}}$ \citep{Chevalier_Fransson06}. Here and throughout this paper we assume a solar abundance for the CSM with mean molecular weight, $\mu=1.4$.

The velocity of the forward shock directly follows as $v_s\approx0.71~R_{FS}/t$ for $s=0$,  and $v_s\approx 0.88~R_{FS}/t$ in the case of $s=2$.  For $M_{\rm ej,Ch}=1$, $E_{\rm K,51}=1$, and fiducial CSM densities ($n_0 = 1$, $A_{\star} = 1$), the forward shock velocity is $0.38\,c$ and $0.17\,c$, respectively, at $t=10$ days since explosion.

We note that this self-similar solution only applies while the reverse shock is interacting with the outer layers of the ejecta distribution characterized by $v \gtrsim v_b$.  The location of the reverse shock during this time trails just behind the contact discontinuity radius, with $R_{RS}/R_c\approx 0.97$ for $s=0$, and $R_{RS}/R_c\approx 0.98$ ($s=2$; \citealt{Chevalier82a}). For the fiducial CSM densities and explosion parameters adopted above, the reverse shock begins to probe the inner, exponentially-distributed material on a timescale of $\sim 5$ years after explosion, and the subsequent evolution of the interaction is modified (see \citealt{Dwarkadas_Chevalier98} for further discussion).  Thus the self-similar model is appropriate during the timescale of our radio observations (i.e. days to one year after explosion, see Tables~\ref{snobs}--\ref{snobs_pan}).

\begin{figure*}[!t]
\vspace{-0.4in}
\centerline{\includegraphics[width=5.0in,angle=0]{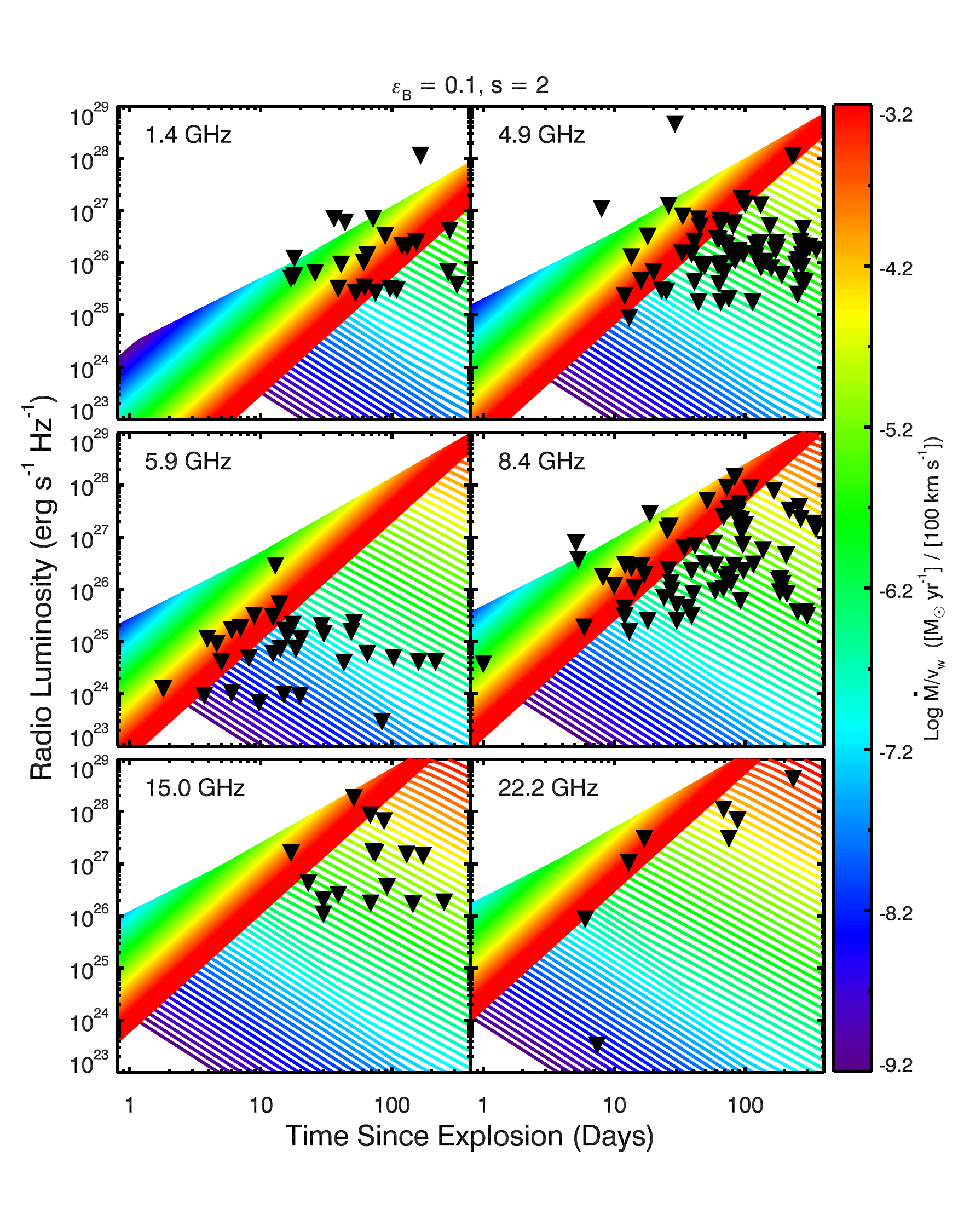}}
\vspace{-0.5in}
\caption{Each panel shows a grid of radio light curve models at a distinct frequency (marked top left), assuming $\epsilon_B = 0.1$, $\epsilon_e = 0.1$, $M_{\rm ej,Ch}=1$, $E_{\rm K,51}=1$, and $s=2$. The model grids span a range of $\dot{M}/v_w$ values ($\dot{M}/v_w = 6\times 10^{-10}-6 \times 10^{-4}~{{{\rm M}_{\odot}\ {\rm yr}^{-1}} \over {{\rm 100\ km\ s}^{-1}}}$) with spacing of 0.1 dex, as described in Section \ref{sec:modcomp}, with purple-blue models corresponding to low $\dot{M}/v_w$ values and red models to high $\dot{M}/v_w$. Observed 3$\sigma$ upper limits at each frequency are overplotted as black triangles.}
\label{fig:gridmoda}
\end{figure*}

\begin{figure*}
\vspace{-0.4in}
\centerline{\includegraphics[width=5.0in,angle=0]{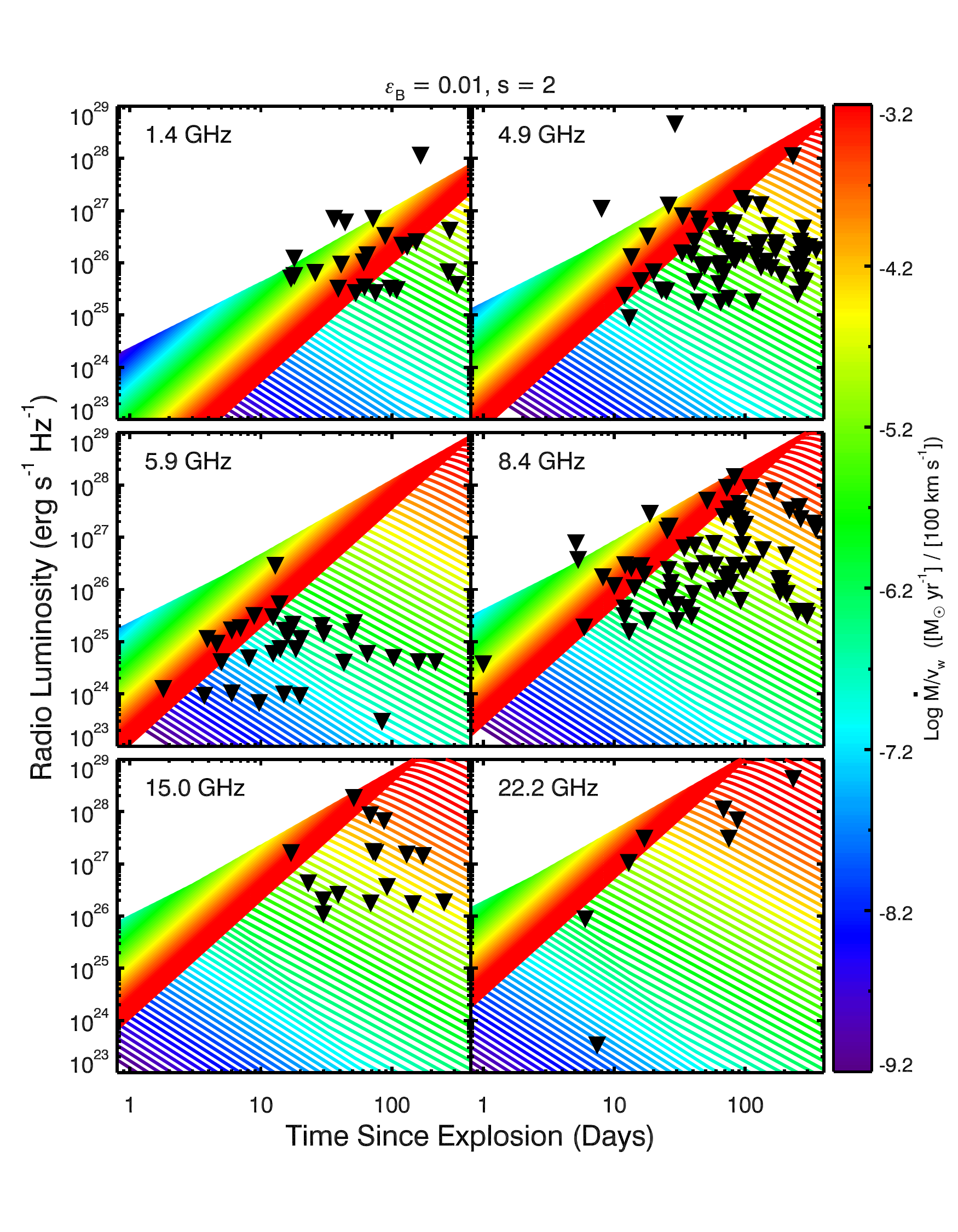}}
\vspace{-0.5in}
\caption{Each panel shows a grids of radio light curve models at a distinct frequency (marked top left), assuming a more conservative $\epsilon_B = 0.01$ (compared to $\epsilon_B = 0.1$ in Figure \ref{fig:gridmoda}) and retaining $\epsilon_e = 0.1$, $M_{\rm ej,Ch}=1$, $E_{\rm K,51}=1$, and $s=2$. The model grid and color scheme represent the same range of $\dot{M}/v_w$ values as in Figure \ref{fig:gridmoda}. Observed 3$\sigma$ upper limits at each frequency are overplotted as black triangles.}
\label{fig:gridmodb}
\end{figure*}

\begin{figure*}
\vspace{-0.4in}
\centerline{\includegraphics[width=5.0in,angle=0]{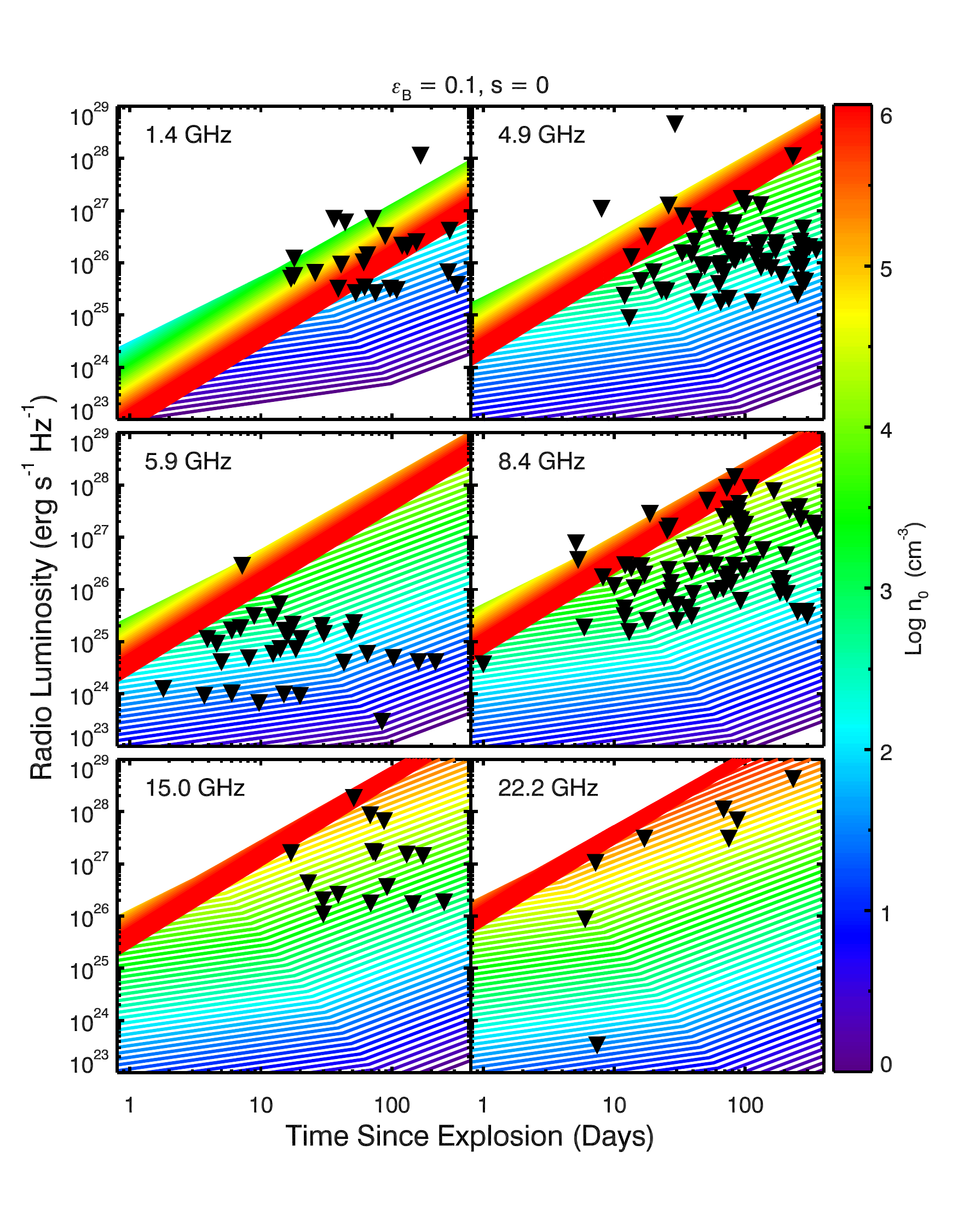}}
\vspace{-0.5in}
\caption{Each panel shows a grids of radio light curve models at a distinct frequency (marked top left), assuming $\epsilon_B = 0.1$, $\epsilon_e = 0.1$, $M_{\rm ej,Ch}=1$, $E_{\rm K,51}=1$, and $s=0$. The model grids span a range of $n_0$ values ($n_0 = 1 - 10^{6}$ cm$^{-3}$) with resolution of 0.1 dex, with purple-blue models corresponding to low $n_0$ values and red models to high $n_0$. Observed 3$\sigma$ upper limits at each frequency are overplotted as black triangles.}
\label{fig:gridmod0a}
\end{figure*}

\begin{figure*}
\vspace{-0.4in}
\centerline{\includegraphics[width=5.0in,angle=0]{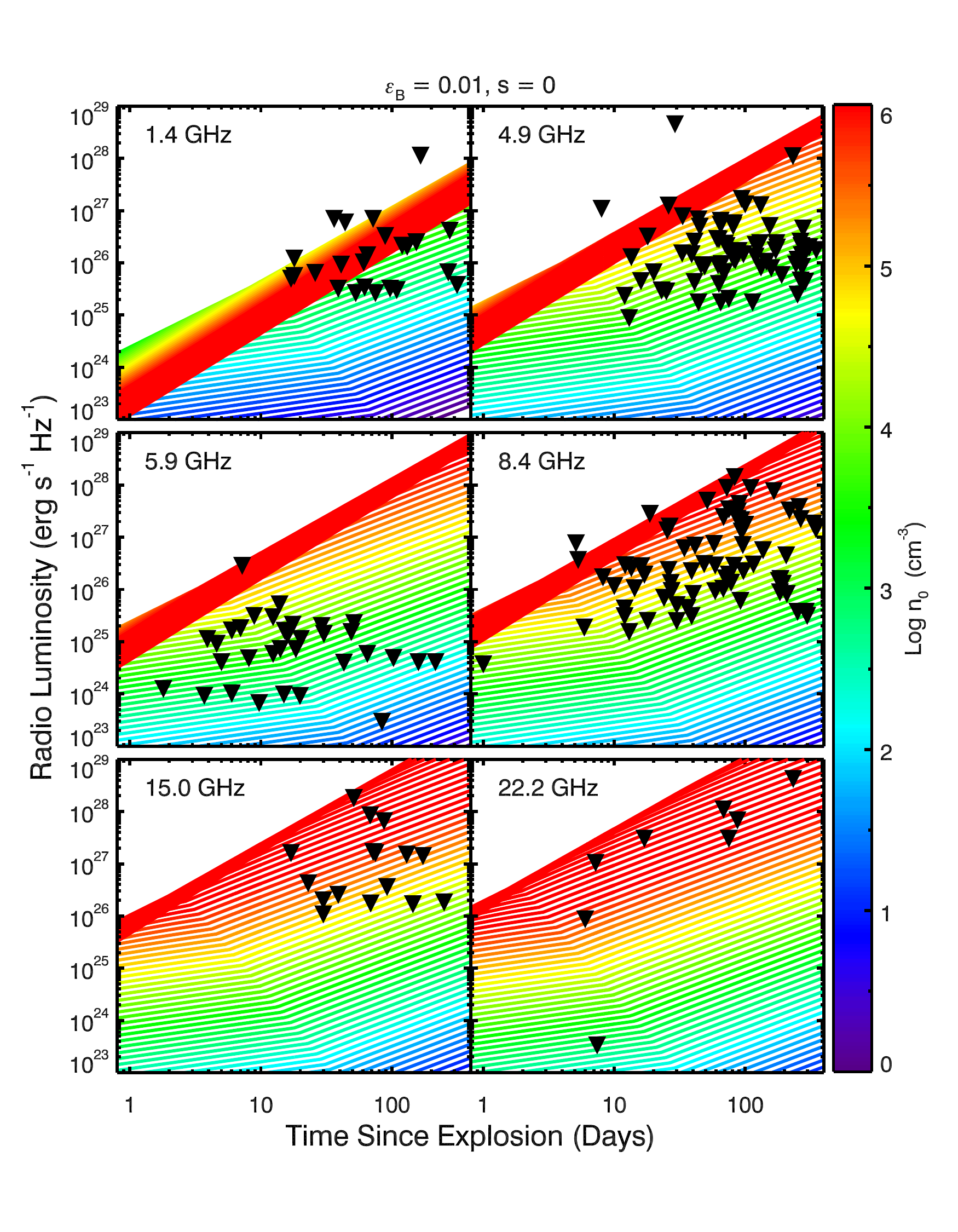}}
\vspace{-0.5in}
\caption{Each panel shows a grids of radio light curve models at a distinct frequency (marked top left), assuming a more conservative $\epsilon_B = 0.01$ (compared to $\epsilon_B = 0.1$ in Figure \ref{fig:gridmod0a}) and retaining $\epsilon_e = 0.1$, $M_{\rm ej,Ch}=1$, $E_{\rm K,51}=1$, and $s=0$. The model grid and color scheme represent the same range of $n_0$ values as in Figure \ref{fig:gridmod0a}. Observed 3$\sigma$ upper limits at each frequency are overplotted as black triangles.}
\label{fig:gridmod0b}
\end{figure*}

\subsection{Magnetic field amplification and acceleration of relativistic particles}
As the forward shock barrels outward into the CSM, electrons are shock-accelerated to relativistic velocities and magnetic fields are amplified through turbulence \citep{Chevalier82b}.  We assume that constant fractions of the post-shock energy density ($\rho_{\rm CSM}~v_s^2$) are shared by the relativistic electrons ($\epsilon_e$), protons ($\epsilon_p$), and amplified magnetic fields ($\epsilon_B$; \citealt{Chevalier98, Chevalier_Fransson06}).  

The strength of the amplified magnetic field in the post-shock region is 
\begin{equation}
B=\sqrt{8\pi\, \epsilon_B\, \rho_{\rm CSM}\, v_s^2}
\label{eqn:bb}
\end{equation}
in cgs units ($B$ in units of G). The magnetic field strength decreases with time and shock radius, because $v_s$ is decreasing---and in the case of wind-stratified CSM, $\rho_{\rm CSM}$ is also decreasing. Combining Equation \ref{eqn:bb} with Equations~\ref{eqn:rs0} and \ref{eqn:rs2}, we find
\begin{equation}
B(s=0)\approx (0.028\ {\rm G})~\epsilon_{B,-1}^{0.5}~n_0^{0.40}~M_{\rm ej,Ch}^{-0.25}~E_{\rm K,51}^{0.35}~t_{10}^{-0.29} 
\end{equation}
\begin{equation}
B(s=2)\approx (1.3\ \rm G)~\epsilon_{B,-1}^{0.5}~A_{\star}^{0.5}~t_{10}^{-1} .
\end{equation}
where $\epsilon_{B,-1}$ is normalized to a fiducial value, $\epsilon_B = 0.1$.

The accelerated electrons populate a high velocity tail that extends beyond the thermal distribution. Above a minimum energy, $E_m\equiv \gamma_m\, m_e\, c^2$ (where $\gamma_m$ is the minimum Lorentz factor of the electrons), the shocked electrons are characterized by a power-law distribution of energies, $N(E > E_m)=N_0\, E^{-p}$.  Observations of other stripped envelope supernovae of Type Ib/c have measured $p=3$ from the optically-thin synchrotron spectral index \citep[e.g.,][]{Weiler_etal86, Berger_etal02, Soderberg_etal05, Soderberg_etal06b}.

In this framework \citep{Chevalier98}, the normalization of the electron distribution is:
\begin{equation}
N_0 = (p-2)~\epsilon_e~\rho_{\rm CSM}~v_s^2~E_m^{p-2}.
\end{equation}
The minimum electron energy is coupled to the shock velocity as \citep{Chevalier_Fransson06}: 
\begin{equation}
E_m=1.2\,  \frac{(p-2)~\epsilon_e~m_p~ v_s^2}{(p-1)~\eta}.
\label{eqn:em}
\end{equation}
where $\eta$ is the compression factor of material in the post-shock region. This equation holds when $E_m >  m_e c^2$; otherwise we set $E_m$ to the electron rest mass energy ($\gamma_m =1$).

Combining Equations~\ref{eqn:rs0}, \ref{eqn:rs2} and \ref{eqn:em}, we derive expressions for the electron minimum Lorentz factor for white dwarf explosions within constant or wind-stratified CSM environments:  
\begin{equation}
\gamma_m(s=0)\approx 32.9~\frac{\epsilon_{e,-1}}{\eta}~\frac{(p-2)}{(p-1)}~ M_{\rm ej,Ch}^{-0.50}~E_{\rm K,51}^{0.70}~n_0^{-0.20}~t_{10}^{-0.58}
\end{equation}
\noindent
\begin{equation}
\gamma_m(s=2)\approx 6.3~\frac{\epsilon_{e,-1}}{\eta}~\frac{(p-2)}{(p-1)}~M_{\rm ej,Ch}^{-0.64}~E_{\rm K,51}^{0.88}~A_{\star}^{-0.24}~t_{10}^{-0.24}
\end{equation}
where $\epsilon_{e.-1}$ is normalized to a fiducial value, $\epsilon_e = 0.1$. For the fiducial values considered here, $\gamma_m$ is larger than unity on timescales of days to weeks after explosion. Henceforth, we assume $\eta = 4$, as for a non-relativistic strong shock. 

The relativistic electrons accelerated to $\gamma_m$ and beyond gyrate in the amplified magnetic field and produce synchrotron emission at a characteristic frequency, $\nu_m=\gamma_m^2 (eB/2\pi m_e c)$.  We find characteristic synchrotron frequency, $\nu_m\sim 10^6$ Hz at $t=10$ days for both the constant density and wind stratified scenarios, and this frequency decays as $\nu_m\propto t^{-1.5}$. Thus $\nu_m$ is well below the observed radio band on the timescale of our observations.  The resultant non-thermal radio emission from the relativistic electron population results in a simple power-law synchrotron spectrum when optically thin, $F_{\nu}\propto \nu^{\beta}$ with $\beta=-(p-1)/2$.

\subsection{Radio luminosity evolution}

Synchrotron self-absorption by emitting electrons within the shock interaction region gives rise to a low-frequency turnover that yields a spectral peak, $\nu_p$, observed in Type Ib/c supernovae \citep{Chevalier98}. In Type I supernovae,  $\nu_p$ is equivalent to the synchrotron self-absorption frequency, since external free-free absorption does not contribute significantly \citep{Chevalier98,Chevalier_Fransson06}. We note that \citet{Panagia_etal06} assume that free-free absorption is the dominant source of opacity in the radio light curves of SNe~Ia, justified by their assumed shock velocity of 10,000 km~s$^{-1}$, which is significantly lower than the thermonuclear SN blast wave velocity derived here. For the fast SN blast waves described by Equations \ref{eqn:rs0} and \ref{eqn:rs2}, the importance of free-free absorption is negligible compared to synchrotron self absorption.

Above $\nu_p$, synchrotron emission is optically thin and the flux density scales as $F_{\nu}\propto \nu^{-1}$ (for $p=3$). Below $\nu_p$,  the emission is optically thick to synchrotron self-absorption, and the spectrum scales as $F_{\nu}\propto \nu^{5/2}$.  As the blast wave expands, the optical depth to synchrotron self-absorption decreases and $\nu_p$ cascades to lower frequencies.  The temporal and spectral evolution of the radio signal is therefore fully determined by the blast wave velocity and the density of the local environment.

Drawing from Equation 1 of \citet{Chevalier98}, we find for $p=3$, the synchrotron flux density at observed frequency $\nu$, is:
\begin{equation}
F_{\nu}\approx 5.31\times 10^{-31}~R_{FS}^2~D^{-2}~B^{-1/2}~\nu^{5/2}  \left[1-e^{-\left(\frac{\nu}{\nu_1}\right)^{-7/2}}\right]
\end{equation}
in cgs units. Here, $D$ is the distance to the supernova and $\nu_1$ is the asymptotic peak frequency joining the optically thick and thin regimes. As the supernova expands and ages, $\nu_1$ decreases as:  
\begin{equation}
\nu_1\approx (4.76\times 10^7~{\rm Hz})~R_{FS}^{2/7}~f^{2/7}~N_0^{2/7}~B^{5/7}.
\end{equation}
defining the frequency at which the optical depth to synchrotron self-absorption is unity. Here, $f$ is the fraction of the spherical volume of the supernova blast which is emitting synchrotron radiation (i.e., $f = 1 - [R_{RS}/R_{FS}]^3$, assuming the radio emission fills the region between the forward shock and the reverse shock). For $s=0$, the solutions of \citet{Chevalier82a} yield $f = 0.38$, while for $s=2$, $f=0.50$. We note that the observed peak frequency ($\nu_p$) is slightly displaced from $\nu_1$ due to exponential smoothing of the two regimes; for $p=3$ this shift is $\nu_p\approx 1.14~\nu_1$. As described in \citet{Chevalier98}, the expressions for $F_{\nu}$ and $\nu_1$ make use of the synchrotron constants applicable for the case of $p=3$ from \citet{Pac70}, specifically $c_1=6.27\times10^{18}$, $c_5=7.52\times 10^{-24}$,  and $c_6=7.97\times 10^{-41}$. 

For low density environments ($n_0 \lesssim 1$, $A_{\star} \lesssim 1$), the effects of synchrotron self-absorption in the centimeter band are minimized such that the optically thin regime is applicable at most times. In this scenario, the synchrotron emission may be simplified as:
\begin{equation}
F_{\nu}\approx 3.95\times 10^{-4}~R_{FS}^3~f~D^{-2}~N_0~B^2~\nu^{-1}~\rm erg~cm^{-2}~s^{-1}~Hz^{-1} .
\end{equation}
Sample light curves are shown in Figure \ref{fig:evla} for $A_{\star} = 0.4$ and $n_0 = 1$ cm$^{-3}$, assuming $\epsilon_e = 0.1$, $\epsilon_B = 0.1$, $M_{\rm ej,Ch}=1$, and $E_{\rm K,51}=1$.

The properties of the CSM density profile and the forward shock evolution together determine the temporal behavior of the optically thin radio flux density.  For $s=0$ and $\gamma_m > 1$, the flux density rises with time as $F_{\nu} \propto t^{0.39}$; once $\gamma_m =1$, the flux density evolution steepens to $F_{\nu} \propto t^{0.97}$ (visible as a kink in the model light curve of Figure \ref{fig:evla} with $n_0 = 1$ cm$^{-3}$ around 100 days after explosion). Meanwhile for $s=2$, the optically-thin scaling is $F_{\nu} \propto t^{-1.60}$ when $\gamma_m > 1$; later, when $\gamma_m$ bottoms out at unity, then $F_{\nu} \propto t^{-1.36}$.  Therefore, the temporal evolution of the optically-thin flux density can be used to estimate the density profile of the CSM.

\subsection{Caveats and Complexities}\label{sec:caveat}

As discussed by \citet{Horesh_etal11, Horesh_etal13}, the most significant uncertainty in converting radio luminosity to a density of the CSM is the efficiency of magnetic field acceleration, $\epsilon_B$. In the case of uniform density CSM and optically-thin synchrotron emission, the density derived from a given flux density will depend on $\epsilon_B$ as $n_0 \propto \epsilon_B^{-0.91}$ when $\gamma_m > 1$, and $n_0 \propto \epsilon_B^{-0.77}$ when $\gamma_m = 1$. For $s=2$ and $\gamma_m > 1$, we find $A_{\star} \propto \epsilon_B^{-0.71}$, and when $\gamma_m$ converges to unity then $A_{\star} \propto \epsilon_B^{-0.61}$. In addition, as seen by comparing Figure \ref{fig:gridmoda} with Figure \ref{fig:gridmodb}, and Figure \ref{fig:gridmod0a} with \ref{fig:gridmod0b}, the synchrotron emission becomes optically thin faster for lower values of $\epsilon_B$.

\begin{figure*}[!t]
\centerline{\includegraphics[width=5.5in,angle=90]{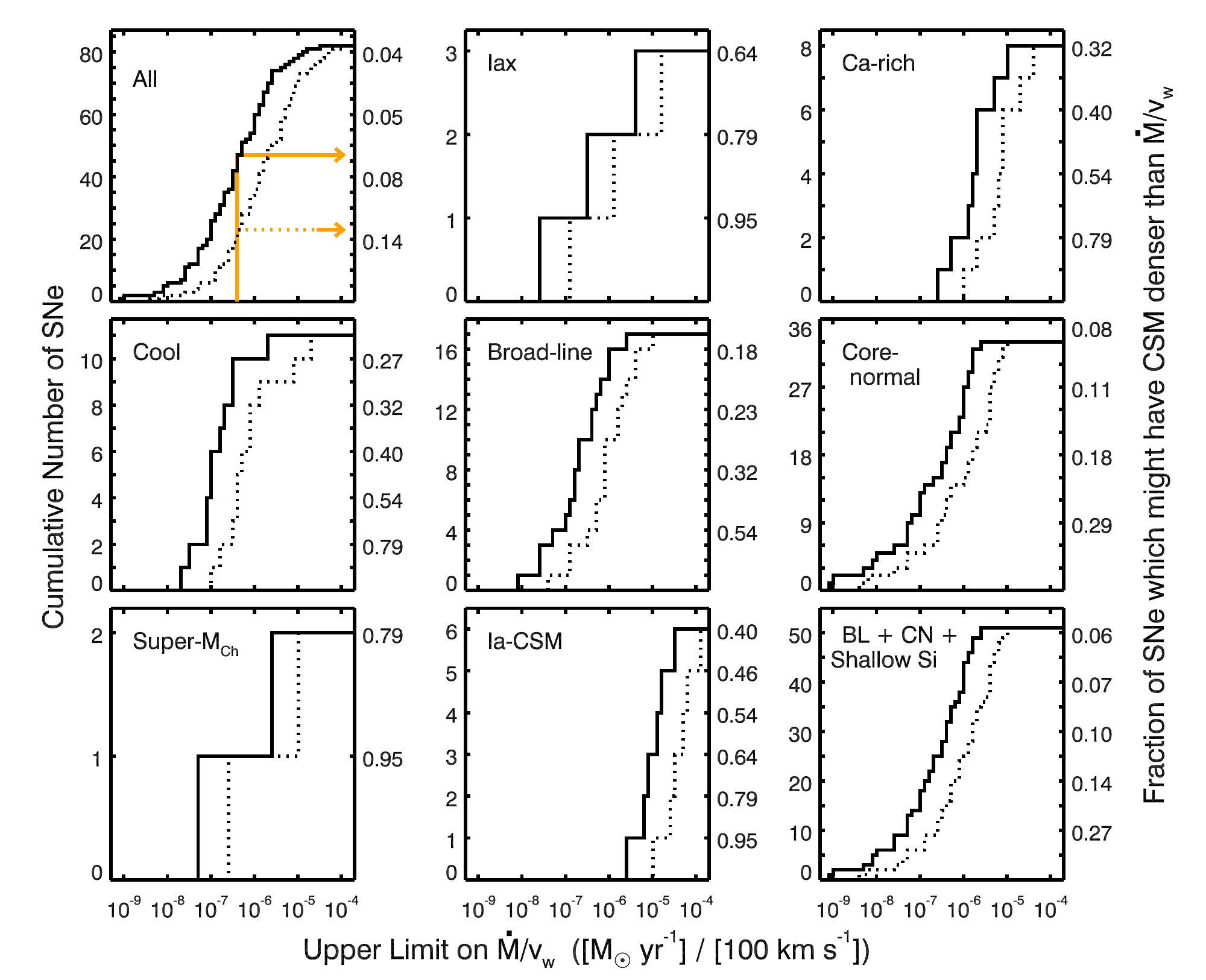}}
\vspace{-0.1in}
\caption{Cumulative histograms of $\dot{M}/v_w$, normalized assuming $v_w$ = 100 km s$^{-1}$. Solid histograms assume $\epsilon_B = 0.1$, while the dotted histograms assume $\epsilon_B = 0.01$; all assume $s=2$, $M_{\rm ej,Ch}=1$, and $E_{\rm K,51}=1$. The top left panel includes all 82 thermonuclear SNe with $\dot{M}/v_w$ constraints, while the other eight panels divide these thermonuclear SNe into the sub-types described in Section \ref{sec:cat}. 02es-like and shallow-silicon supernovae are not shown in individual panels, as there is only one object in our sample of each of these sub-types. The bottom-right panel combines the sub-types of SNe Ia that are cosmologically useful (broad-line, core-normal, and shallow-silicon). The labels on the right side of the plot show what fraction of SNe in the sample might have surroundings denser than $\dot{M}/v_w$, based off simple binomial statistics at 2$\sigma$ significance. The orange lines in the top left panel illustrate that we constrain 47 thermonuclear SNe ($>$94\% of all thermonuclear SNe) to have $\dot{M}/v_w < 4 \times10^{-7}~{{{\rm M}_{\odot}\ {\rm yr}^{-1}} \over {{\rm 100\ km\ s}^{-1}}}$ assuming $\epsilon_B = 0.1$ (orange solid line), and 23 thermonuclear SNe ($>$87\% of all thermonuclear SNe) to meet this criterion if $\epsilon_B = 0.01$ (orange dotted line).}
\label{fig:astarhist}
\end{figure*}

\begin{figure*}
\centerline{\includegraphics[width=5.5in,angle=90]{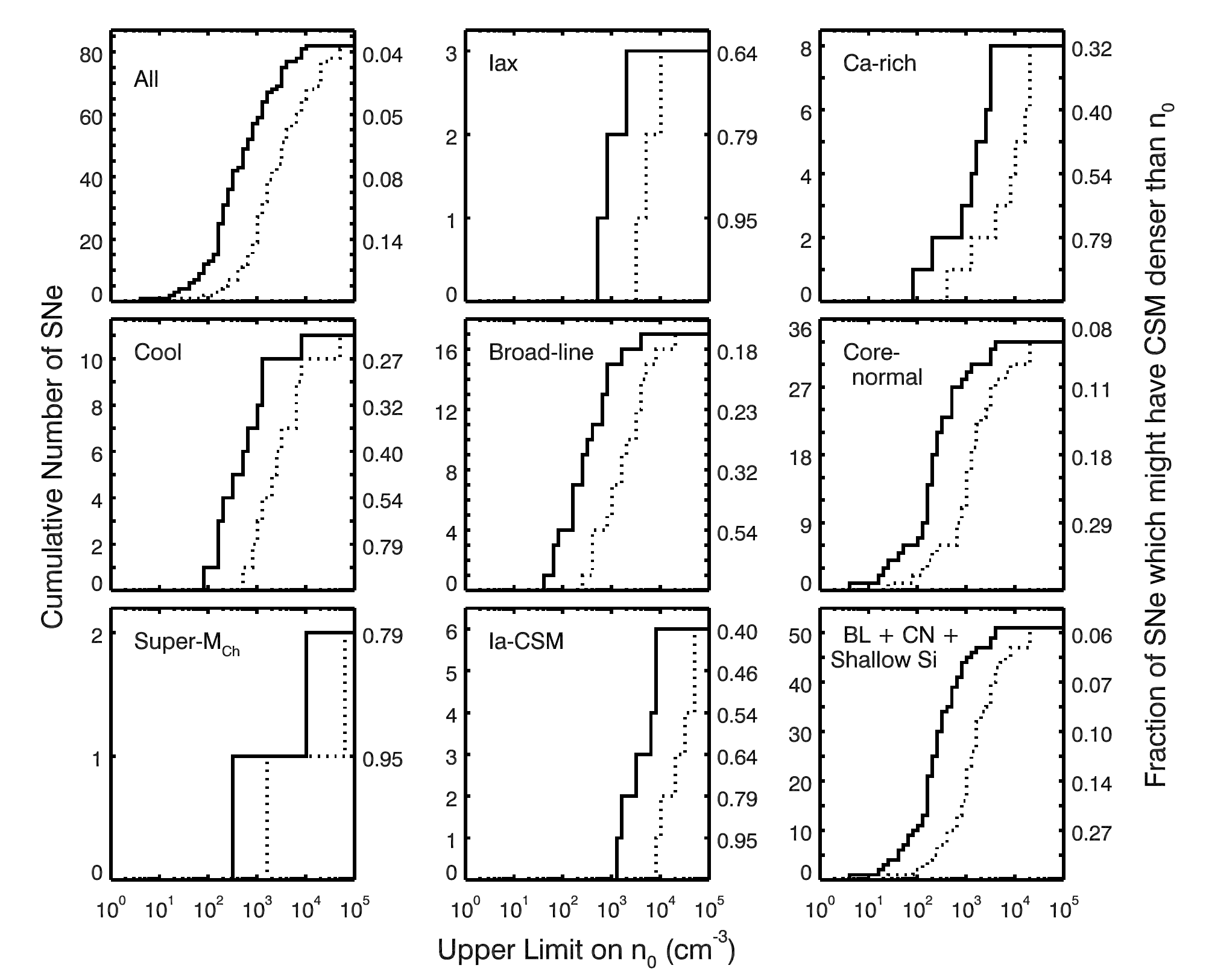}}
\vspace{-0.1in}
\caption{Cumulative histograms of $n_0$, assuming the CSM is uniform in density. Solid histograms assume $\epsilon_B = 0.1$, while the dotted histograms assume $\epsilon_B = 0.01$; all assume $s=0$, $M_{\rm ej,Ch}=1$, and $E_{\rm K,51}=1$. The top left panel includes all 82 thermonuclear SNe with $n_0$ constraints, while the other eight panels divide these thermonuclear SNe into the sub-types described in Section \ref{sec:cat}. 02es-like and shallow-silicon supernovae are not shown in individual panels, as there is only one object in our sample of each of these sub-types. The bottom-right panel combines the sub-types of SNe Ia that are cosmologically useful (broad-line, core-normal, and shallow-silicon). The labels on the right side of the plot show what fraction of SNe in the sample might have surroundings denser than $n_0$, based off simple binomial statistics at 2$\sigma$ significance.}
\label{fig:n0hist}
\end{figure*}

Radio light curve models are also affected by the assumed values of SN ejecta mass and explosion energy. If we take conservatively low values of $M_{\rm ej,Ch}=0.1$ and $E_{\rm K,51}=0.1$, we find forward shock velocities of $0.13\,c$ and $0.29\,c$ at $t=10$ days since explosion for $A_{\star} = 1$ and  $n_0 = 1$ respectively. These blast wave velocities are high enough that synchrotron self absorption still dominates over free-free absorption. In the case of uniform density CSM and optically-thin synchrotron emission, the density derived from a given flux density will depend on ejecta mass and kinetic energy as $n_0 \propto M_{\rm ej,Ch}^{2.05}\ E_{\rm K,51}^{-2.86}$ when $\gamma_m > 1$, and $n_0 \propto M_{\rm ej,Ch}^{1.35}\ E_{\rm K,51}^{-1.88}$ when $\gamma_m = 1$. For $s=2$ and $\gamma_m > 1$, we find $A_{\star} \propto M_{\rm ej,Ch}^{1.14}\ E_{\rm K,51}^{-1.57}$, and when $\gamma_m$ converges to unity then $A_{\star} \propto M_{\rm ej,Ch}^{0.59}\ E_{\rm K,51}^{-0.80}$. While these dependencies are quite strong, uncertainties in $M_{\rm ej,Ch}$ and $E_{\rm K,51}$ do not affect our constraints on $n_0$ and $A_{\star}$ as dramatically as they may first appear, because e.g., in low-luminosity thermonuclear explosions,  both $M_{\rm ej,Ch}$ and $E_{\rm K,51}$ are suppressed (compared to fiducial SN~Ia values). While decreasing the assumed $E_{\rm K,51}$ yields a less stringent constraint on $\rho_{\rm CSM}$, decreasing the assumed $M_{\rm ej,Ch}$ implies a stronger constraint, and therefore co-varying changes to $M_{\rm ej,Ch}$ and $E_{\rm K,51}$ largely counteract one another. The effects of assumed $M_{\rm ej,Ch}$ and $E_{\rm K,51}$ will be explored in more detail in \S \ref{sec:indiv}.
 
 The predicted radio emission is modified if cooling of the electrons steepens the spectrum, thus decreasing the emission as considered by \citet{Chevalier_Fransson06} for the case of SNe Ib/c. The case of synchrotron losses is analogous for the two supernova types and Equation 11 of \citet{Chevalier_Fransson06} shows that synchrotron losses are unimportant in both cases. For Inverse Compton losses, thermonuclear SNe have a higher peak optical luminosity, $\sim 10^{43}$ erg s$^{-1}$, than typical SNe Ib/c, but the lower ejecta mass leads to a larger distance of the relativistic electrons from the photosphere, so Inverse Compton losses are only of marginal importance, if any, at the time of optical maximum light.

\section{Comparing Observations with Radio Light Curve Models}
\label{sec:modcomp}

\subsection{Upper Limits on CSM Density}
The goal of the modelling described above is to translate our limits on the radio luminosity of thermonuclear SNe to constraints on the density of the CSM around thermonuclear SNe. We therefore calculated grids of radio light curves at all observed frequencies using these models, for comparison with observations. 

For a wind CSM ($s=2$), we sample $\dot{M}/v_w$ space logarithmically with a resolution of 0.1 dex. The model grid spans mass loss rates $\dot{M}/v_w = 6\times 10^{-10}-6 \times 10^{-4}~{{{\rm M}_{\odot}\ {\rm yr}^{-1}} \over {{\rm 100\ km\ s}^{-1}}}$, assuming $M_{\rm ej} = M_{\rm ej,Ch}$, $E_{\rm K,51} = 1$, and $\epsilon_e = 0.1$.

The grid of wind CSM models at the most commonly observed frequencies is shown in Figures \ref{fig:gridmoda} and \ref{fig:gridmodb}, with radio limits overplotted. The two figures are for different values of $\epsilon_B$. In Tables \ref{snobs}--\ref{snobs_pan}, we also list the upper limits on $\dot{M}/v_w$, normalized to $v_w = 100$ km s$^{-1}$ and $\epsilon_B = 0.1$. In the few cases where radio limits lie above the optically-thick locus of the model grid, they do not constrain the density of the CSM at all, and in these cases ``N/A" is listed for the limit on  $\dot{M}/v_w$. 

At first glance, our constraints on $\dot{M}/v_w$ appear similar to those listed by \citet{Panagia_etal06} in their Table 3. However, they assume $v_w$ = 10 km s$^{-1}$, while our values of $\dot{M}/v_w$ are normalized to 100 km s$^{-1}$. Therefore, the constraints on $\dot{M}/v_w$ from the Panagia et al.\ model are a factor of $11.1\pm6.2$ less constraining than the limits derived from our model. In addition, \citet{Panagia_etal06} quote 2$\sigma$ constraints on $\dot{M}/v_w$, while those presented here are 3$\sigma$, so the discrepancy in $\dot{M}/v_w$ is in fact somewhat larger than a factor of 11. Most of this discrepancy is attributable to Panagia et al.'s assumption of a slower blast wave ($v_s$ = 10,000 km s$^{-1}$); different assumptions about synchrotron opacity also affect the calculation. We note that \cite{Panagia_etal06} do not model their radio limits with uniform CSM.

Using our model of uniform-density CSM ($s=0$), we sample $n_0$ space logarithmically with a resolution of 0.1 dex, from $n_0 = 1$ cm$^{-3}$ to $n_0 = 10^6$ cm$^{-3}$. As for $s=2$, we assume $M_{\rm ej,Ch} = 1$, $E_{\rm K,51} = 1$, and $\epsilon_e = 0.1$. These models are compared with our upper limits in Figures \ref{fig:gridmod0a} and \ref{fig:gridmod0b}, for $\epsilon_B$ = 0.1 and 0.01, respectively. Upper limits on $n_0$ are also listed in Tables \ref{snobs}--\ref{snobs_pan}, assuming $\epsilon_B = 0.1$ (again, ``N/A" is listed if a limit sits above the optically-thick model locus).

It is clear that later observations, several years after explosion, will be significantly more constraining upon uniform density surroundings. Such later observations will be presented and discussed in a forthcoming publication. On the other hand, early observations are most constraining for a wind CSM.

It should also be noted that the assumption made in this paper, that the inner SN ejecta remain unshocked, is only valid out to time scales of one year if $n_0 \lesssim$ 100 cm$^{-3}$ or $\dot{M}/v_w = 10^{-5}~{{{\rm M}_{\odot}\ {\rm yr}^{-1}} \over {{\rm 100\ km\ s}^{-1}}}$. Our treatment of the highest CSM densities considered here is only valid at earlier times (out to Day 40 for $\dot{M}/v_w = 10^{-4}~{{{\rm M}_{\odot}\ {\rm yr}^{-1}} \over {{\rm 100\ km\ s}^{-1}}}$, or Day 162 for $n_0 = 10^3$ cm$^{-3}$). SNe expanding into such dense CSM will also likely suffer free-free absorption. We mark observations that can not be completely described by the model of outer ejecta presented here---because the luminosity limit is not terribly deep or because the observation was taken at a relatively late time---with a superscript in Tables \ref{snobs_old} and \ref{snobs_pan} (this situation does not apply to any of the observations in Table \ref{snobs}). Proper modeling of these highest CSM densities are outside the scope of this publication, but should be addressed in the future.

Cumulative histograms of $\dot{M}/v_w$ are shown in Figure \ref{fig:astarhist} (normalized to $v_w = 100$ km~s$^{-1}$), for both our entire sample and broken up by thermonuclear SN sub-type. On the right side of each panel, we estimate the fraction of thermonuclear SNe that could have CSM denser than the plotted equivalent $\dot{M}/v_w$, and still be consistent with our measured limits at 2$\sigma$ significance. For example, take the case where we have a sample of 47 thermonuclear SNe which have all been constrained to have $\dot{M}/v_w < 4 \times10^{-7}~{{{\rm M}_{\odot}\ {\rm yr}^{-1}} \over {{\rm 100\ km\ s}^{-1}}}$ (the average $\dot{M}/v_w$ for Galactic symbiotic binaries; see \S \ref{sec:symb}; Figure \ref{fig:astarhist}). There is a 95.5\% (2$\sigma$) binomial probability of this occurrence if the probability of success on a single trial is 0.064. In other words, we estimate that $<$6.4\% of thermonuclear SNe have $\dot{M}/v_w > 4 \times10^{-7}~{{{\rm M}_{\odot}\ {\rm yr}^{-1}} \over {{\rm 100\ km\ s}^{-1}}}$. This simple analysis assumes that our observed thermonuclear SNe are representative of thermonuclear SNe as a whole (or of the sub-types, for the histograms broken down by type).

Similar cumulative histograms are shown in Figure \ref{fig:n0hist} for uniform-density CSM. These limits are less constraining on progenitor models; we find that 13 SNe ($>$79\% of all thermonuclear SNe at 2$\sigma$ significance) constrain $n_0 < 100$ cm$^{-3}$ for $\epsilon_B = 0.1$. As previously discussed, future work presenting later radio observations of SNe Ia (1--100 years after explosion) will be more constraining on $n_0$.

\subsection{Implications for Symbiotic Progenitors} \label{sec:symb}

The early-time limits on the radio luminosity presented here place significant constraints on symbiotic progenitors for thermonuclear SNe. Specifically, our limits constrain progenitors with red giant companions undergoing steady mass loss immediately preceding the thermonuclear SN. If there is a significant delay between the giant mass loss and the SN \citep{Justham11, DiStefano_etal11}, or if a cavity is excavated in the circumbinary material by e.g., an accretion-powered wind \citep{Hachisu_etal99}, the CSM would not be expected to follow the simple $\rho \propto r^{-2}$ we consider here, and would be better constrained by observations longer after SN explosion.

\begin{figure}
\centerline{\includegraphics[width=3.5in,angle=90]{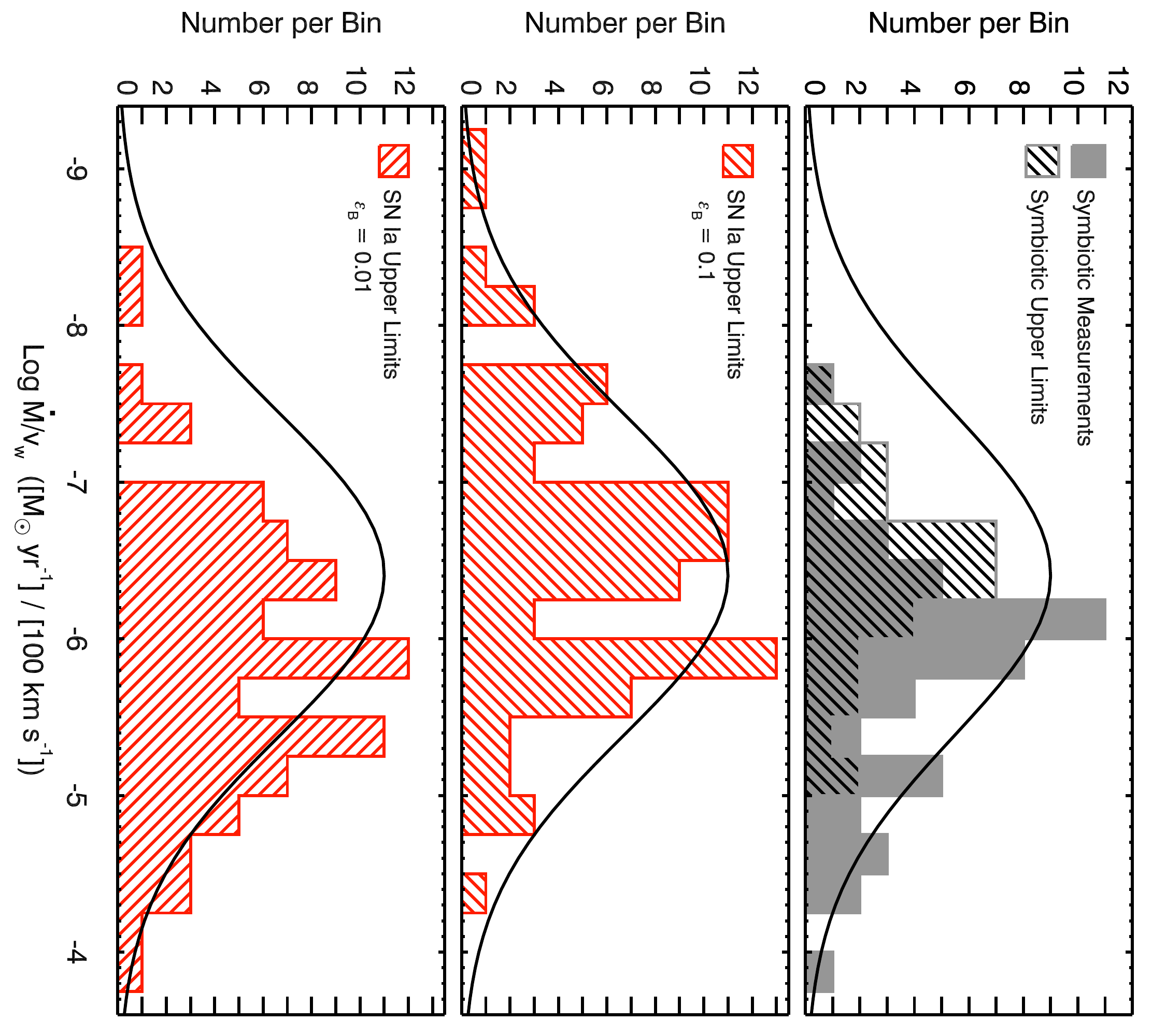}}
\vspace{-0.1in}
\caption{Histograms comparing measurements and constraints on $\dot{M}/v_w$ for symbiotic binaries with upper limits for thermonuclear SNe, normalized to $v_w$ = 100 km s$^{-1}$. The top panel shows data for symbiotic binaries from \citet{Seaquist_etal93}, with measurements as the gray shaded histogram and upper limits as the hatched histogram. The best-fit log-normal distribution to these data is shown as a black solid line. The middle panel shows the distribution of $\dot{M}/v_w$ upper limits for thermonuclear SNe as the red hatched histogram, assuming $M_{\rm ej,Ch}=1$, $E_{\rm K,51}=1$, and $\epsilon_B = 0.1$; the fitted log-normal distribution for symbiotics is overplotted as a black solid line. The bottom panel is identical to the middle panel, except $\epsilon_B = 0.01$ is assumed in the calculation of $\dot{M}/v_w$. We note that the amplitude of the log-normal fit to the Galactic symbiotics is arbitrary, and adjusted for diagnostic purposes.}
\label{fig:symbhist}
\end{figure}

Radio observations of Galactic symbiotic stars yield robust measurements of $\dot{M}/v_{w}$ in these systems, probing the ionized wind CSM via thermal free-free emission \citep{Seaquist_etal84, Seaquist_Taylor90, Seaquist_etal93}. The distribution of $\dot{M}/v_w$ for symbiotic binaries is shown in Figure \ref{fig:symbhist}, presenting the catalog of \citet{Seaquist_etal93}. Seaquist et al.\ note that their catalog contains ``essentially every known symbiotic star observable with $\delta > -45^{\circ}$" as of 1993; the sample spans both S-type and D-type systems, and a range of IR and ultraviolet properties. 

When a symbiotic system is detected at radio frequencies (filled gray histogram), its $\dot{M}/v_w$ was estimated by \citet{Seaquist_etal93} assuming a fully-ionized, spherically-symmetric wind CSM and the equations of \citet{Wright_Barlow75}. In other cases, when a symbiotic binary is not detected, a 3$\sigma$ upper limit on its $\dot{M}/v_w$ was given by \citet{Seaquist_etal93}, and is plotted in the top panel of Figure \ref{fig:symbhist} as the hatched histogram. Note that the radio measurements of symbiotic stars constrain $\dot{M}/v_{w}$, as we also do for thermonuclear SNe here. While \citet{Seaquist_etal93} assume $v_w$ = 30 km s$^{-1}$ in their analysis, in our Figure \ref{fig:symbhist} we have normalized their reported $\dot{M}/v_w$ to $v_w$ = 100 km s$^{-1}$, for consistency with the rest of our study. \citet{Seaquist_etal93} also note that their measurements of symbiotic $\dot{M}/v_w$ should be viewed as lower limits, because the red giant wind is only partially ionized in most systems. However, the total neutral + ionized $\dot{M}/v_w$ is probably within a factor of $\sim$2 of the estimates presented by \citet{Seaquist_etal93} and plotted in Figure \ref{fig:symbhist}.

We used these measurements to estimate the distribution of $\dot{M}/v_w$ for the known symbiotic systems. To do this, we used a Bayesian Markov Chain Monte Carlo technique to fit a log-normal distribution to both the measured $\dot{M}/v_w$ values and the censored data (upper limits on $\dot{M}/v_w$), assuming common uncertainties of 0.3 dex and flat priors. We find that the distribution of log($\dot{M}/v_w$) for Galactic symbiotic binaries has a best-fit model with a mean, $\mu = -6.41^{+0.14}_{-0.13}$ and standard deviation, $\sigma = 1.03^{+0.13}_{-0.11}$ (top panel of Figure \ref{fig:symbhist}).

Our next step was to compare the collection of upper limits for the themonuclear SNe to the best-fit distribution for symbiotic binaries (middle and lower panels of Figure \ref{fig:symbhist}). To do this, we make the simplifying assumption that there are two populations of thermonuclear SNe: those with symbiotic progenitors that behave like the known symbiotics, and a second population of unspecific origin that does not produce radio emission detectable with current facilities. We generate Monte Carlo samples with a varying fraction $f$ drawn from the symbiotic population, then censor these data with the observed thermonuclear SN limits. We take the test statistic as the fraction of trials in which no SNe Ia would be detected (as observed).

For $\epsilon_B = 0.1$, at the level of $p = 0.05$, the maximum fraction of symbiotic progenitors allowable by our radio limits is $f = 0.07$. At a more conservative $p=0.01$, this fraction is $f=0.10$. The corresponding $p= 0.05$ and $p= 0.01$ values for $\epsilon_B = 0.01$ are $f = 0.10$ and 0.16, respectively. Thus we conclude that no more than $\sim$10--15\% of thermonuclear SNe have symbiotic progenitors comparable to Galactic binaries. A relatively high rate of such progenitors ($\sim25$\%), as inferred from other studies \citep{Sternberg_etal11, Maguire_etal13}, is inconsistent with our observations to a high degree of confidence. 

Since this analysis is based on the population properties of SNe Ia, and the distribution of symbiotic mass loss rates is relatively broad, these constraints can continue to be improved with additional radio continuum observations. Since symbiotic progenitors appear to be relatively uncommon, a larger set of observations of modest depth would improve the constraints more than a smaller number of more sensitive observations or a focus on the nearest SNe.

\section{Notes on Some Individual Sub-types and Sources}
\label{sec:indiv}

\subsection{SNe~Iax}

\begin{figure}
\centerline{\includegraphics[width=2.9in,angle=90]{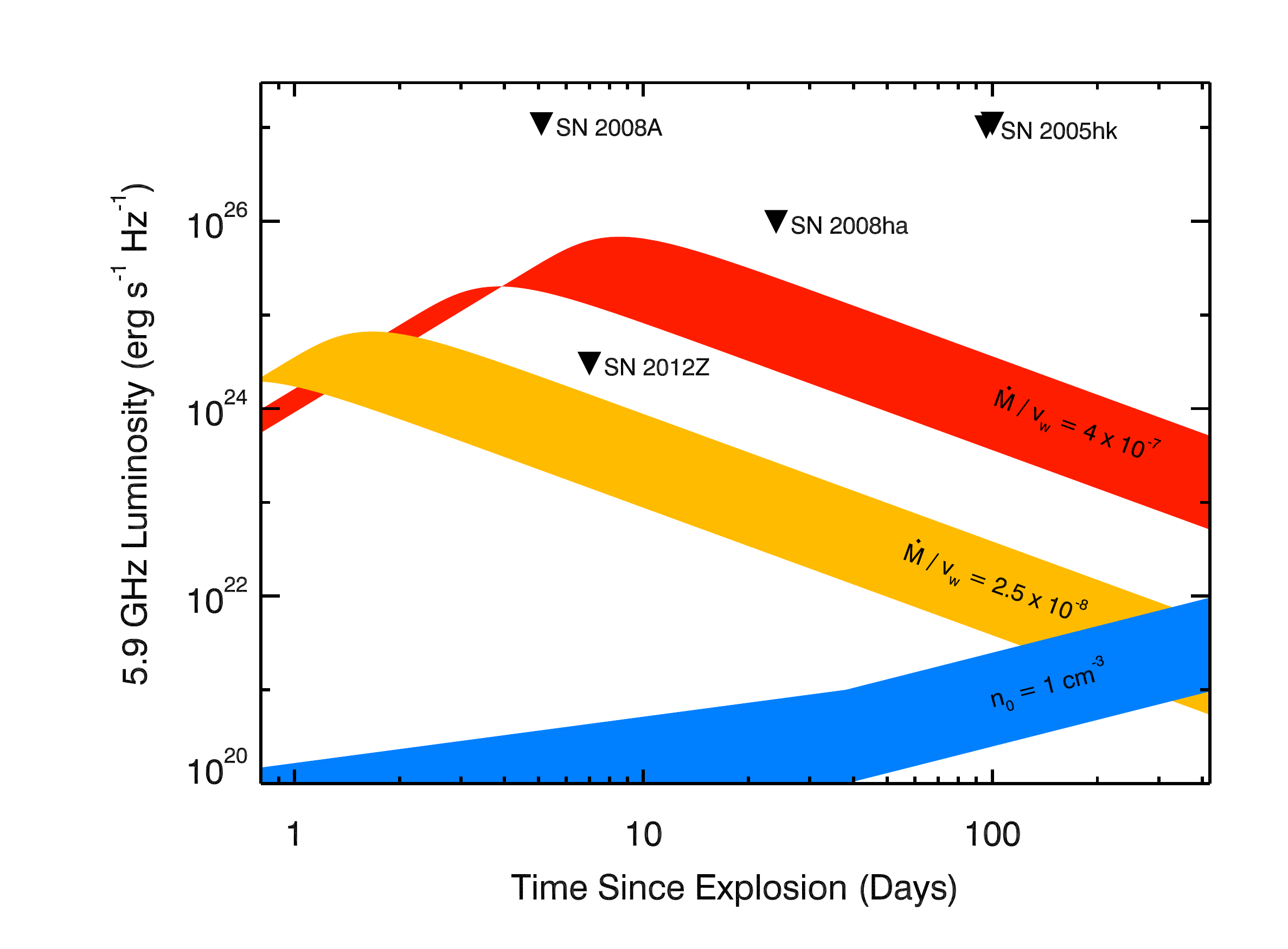}}
\vspace{-0.1in}
\caption{5.9 GHz upper limits for the four SNe~Iax in our sample (downward facing black triangles), compared with models for a thermonuclear SN interacting with CSM, assuming SN~Iax-appropriate $M_{\rm ej,Ch}=0.4$ and $E_{\rm K,51}=0.1$. Observations that were carried out at frequencies differing from 5.9 GHz were normalized to 5.9 GHz assuming an optically-thin synchrotron spectral index, $\alpha = -1$. The red model light curve is for a wind-stratified CSM with $\dot{M}/v_w = 4\times 10^{-7}~{{{\rm M}_{\odot}\ {\rm yr}^{-1}} \over {{\rm 100\ km\ s}^{-1}}}$ (the average value for Galactic symbiotics). The orange model light curve corresponds to $\dot{M}/v_w = 2.5\times 10^{-8}~{{{\rm M}_{\odot}\ {\rm yr}^{-1}} \over {{\rm 100\ km\ s}^{-1}}}$ (as might be expected for a He star with a relatively strong wind; corresponding to $\dot{M} = 1\times 10^{-7}~{{{\rm M}_{\odot}\ {\rm yr}^{-1}}}$ for $v_w = 400$ km s$^{-1}$). The blue light curve is for uniform density surroundings with $n_0 = 1$ cm$^{-3}$. Each model light curve is actually a band of finite width, illustrating the spread in light curves produced by $\epsilon_B$ values ranging $0.01-0.1$; $\epsilon_e = 0.1$ is always assumed.}
\label{fig:iax}
\vspace{0.1in}
\end{figure}

Our constraints on $\rho_{\rm CSM}$ for SNe~Iax assuming $M_{\rm ej,Ch}=1$ and $E_{\rm K,51}=1$ can be found in Figures \ref{fig:astarhist} and \ref{fig:n0hist}, along with Tables \ref{snobs} and \ref{snobs_old}. However, these low-luminosity explosions likely have significantly lower ejecta mass and explosion kinetic energy. Reasonable assumptions for the class are $M_{\rm ej,Ch} \approx 0.4$ ($\sim$0.5 M$_{\odot}$) and $E_{\rm K,51} \approx 0.1$ \citep{Foley_etal13}; detailed calculations on SN\,2008ha imply even lower ejecta masses and energies in some cases \citep{Foley_etal09, Foley_etal10, Valenti_etal09}. SNe Iax and appropriately-scaled light curve models are shown in Figure \ref{fig:iax}.

Based on the pre-explosion detection of SN~Iax 2012Z and claims that its companion may be a He star \citep{Foley_etal13, McCully_etal14}, the CSM of He stars is worth considering. He stars have been observed to have winds with $\dot{M} = 10^{-11}-10^{-7}$ M$_{\odot}$ yr$^{-1}$ and $v_w$ = few hundred -- few thousand km s$^{-1}$ \citep{Jeffery_Hamann10}. Our limits from the pre-upgrade VLA are far from constraining of He star companions; we find $\dot{M} \lesssim {\rm few} \times 10^{-6}-10^{-4}$ M$_{\odot}$ yr$^{-1}$  for $v_w$ = 400 km s$^{-1}$, $M_{\rm ej,Ch}=0.4$, and $E_{\rm K,51}=0.1$. However, our Jansky VLA limit on SN\,2012Z approaches the He star parameter space ($\dot{M} < 2 \times 10^{-7}$ M$_{\odot}$ yr$^{-1}$ for $v_w$ = 400 km s$^{-1}$, $M_{\rm ej,Ch}=0.36$, and $E_{\rm K,51}=0.1$). Future nearby SNe~Iax are ideal targets for deep radio observations.

\subsection{Ca-rich SNe}

\begin{figure}
\centerline{\includegraphics[width=2.9in,angle=90]{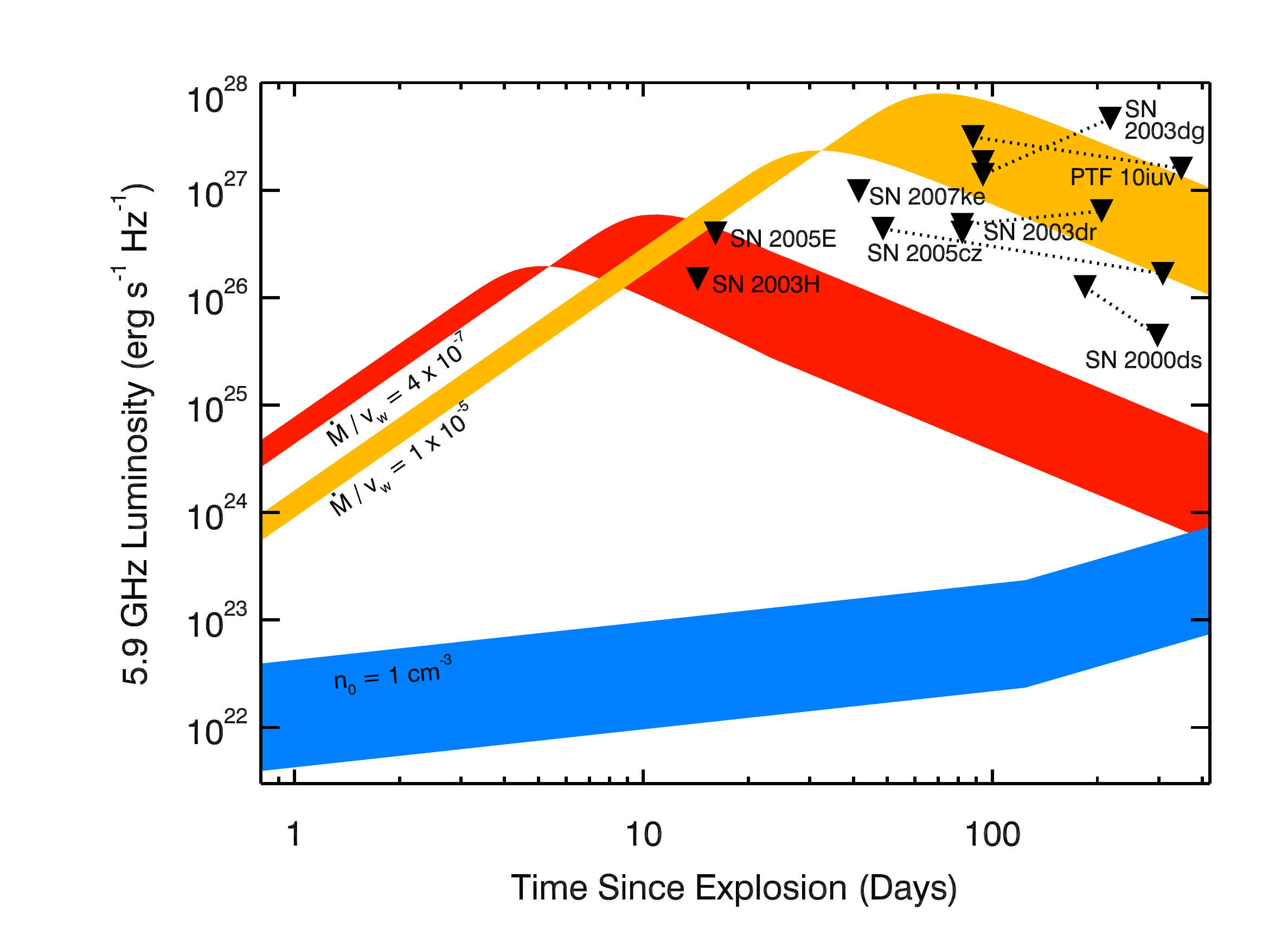}}
\vspace{-0.1in}
\caption{5.9 GHz upper limits for the eight Ca-rich SNe in our sample (downward facing black triangles), compared with models for a thermonuclear SN interacting with CSM, assuming explosion parameters appropriate to Ca-rich explosions ($M_{\rm ej,Ch}=0.2$ and $E_{\rm K,51}=0.4$). Observations that were carried out at frequencies differing from 5.9 GHz were normalized to 5.9 GHz assuming an optically-thin synchrotron spectral index, $\alpha = -1$. SNe with observations at more than one epoch have their limits connected by black dotted lines. The red model light curve is for a wind-stratified CSM with $\dot{M}/v_w = 4\times 10^{-7}~{{{\rm M}_{\odot}\ {\rm yr}^{-1}} \over {{\rm 100\ km\ s}^{-1}}}$ (the average value for Galactic symbiotics). The orange model light curve corresponds to $\dot{M}/v_w = 1\times 10^{-5}~{{{\rm M}_{\odot}\ {\rm yr}^{-1}} \over {{\rm 100\ km\ s}^{-1}}}$ (as might be expected for tidal tails from a disrupted white dwarf). The blue light curve is for uniform density surroundings with $n_0 = 1$ cm$^{-3}$. Each model light curve is actually a band of finite width, illustrating the spread in light curves produced by $\epsilon_B$ values ranging $0.01-0.1$; $\epsilon_e = 0.1$ is always assumed.}
\label{fig:carich}
\vspace{0.1in}
\end{figure}

Like SNe~Iax, the low-luminosity Ca-rich SNe have lower ejecta masses and explosion kinetic energies that standard SNe~Ia. Here we take the findings of\citet{Perets_etal10}: $M_{\rm ej,Ch} \approx 0.2$ ($\sim$0.3 M$_{\odot}$) and $E_{\rm K,51} \approx 0.4$. Limits on Ca-rich SNe and appropriate light curve models are plotted in Figure \ref{fig:carich}. Current data, mostly from the pre-upgrade VLA, constrains $\dot{M}/v_w \lesssim 2\times 10^{-7}~{{{\rm M}_{\odot}\ {\rm yr}^{-1}} \over {{\rm 100\ km\ s}^{-1}}}$ for SN\,2003H and $\dot{M}/v_w \lesssim 10^{-5}~{{{\rm M}_{\odot}\ {\rm yr}^{-1}} \over {{\rm 100\ km\ s}^{-1}}}$ for most objects in the class, assuming ejecta mass and explosion energy appropriate to Ca-rich explosions.

\citet{Perets_etal10} hypothesize that Ca-rich transients are due to He detonations on white dwarfs, due to accretion of material from a He white dwarf. We can rule out the presence of tidal tails around Ca-rich SNe, in the case where stripping of the He white dwarf occurred a few years before SN explosion--- \citet{Raskin_Kasen13} predict the CSM density would correspond to  $\dot{M} = 10^{-5}-10^{-2}$ M$_{\odot}$ yr$^{-1}$. We can also rule out the presence of strong accretion-powered outflows or winds, as might be expected if a disrupted white dwarf is accreted, if the winds were powered for several years preceding the Ca-rich explosion.

\subsection{Cosmologically useful SNe Ia: Broad-line, Core-normal, and Shallow-Silicon}

It is the core-normal, broad-line, and shallow-silicon SNe Ia which are most often exploited as standardizeable candles, and which are well described by $M_{\rm ej,Ch}=1$ and $E_{\rm K,51}=1$. If we perform a similar statistical analysis as that described in \S \ref{sec:symb} on the 51 SNe belonging to these sub-types, we find that, for $\epsilon_B = 0.1$, the maximum fraction of symbiotic progenitors allowable by our radio limits is $f = 0.10$ at significance $p=0.05$, or $f =0.15$ for $p =0.01$. The corresponding $p= 0.05$ and $p= 0.01$ values for $\epsilon_B = 0.01$ are respectively  $f = 0.14$ and 0.21. 

We therefore arrive at a similar conclusion as searches for early-time excesses in optical light curves of thermonuclear SNe, but using a completely independent technique. From limits on early-time excesses in $\sim$100 optical light curves, \citet{Bianco_etal11} find that Roche-lobe-filling red giants must comprise $<$20\% of SN~Ia companions, at 3$\sigma$ significance. We note that our radio limits apply to red giant + white dwarf binaries with much larger separations than the Roche-lobe-filling binary assumed in the models of \citet{Kasen10}. Our constraints on symbiotic progenitors apply to wind-fed or wind-Roche-lobe-overflow systems \citep{Mohamed_Podsiadlowski07,Mikolajewska12}, being modeled on the observed Galactic population of symbiotic binaries.

There have been claims in the literature that there are differences in CSM between broad-line and core-normal SNe~Ia, with broad-line SNe~Ia more likely to display blueshifted \ion{Na}{1} D absorption \citep{Foley_etal12, Maguire_etal13}. Our radio non-detections of SNe~Ia of both sub-types do not support, but can not disprove, this claim.

\subsection{Super-Chandrasekhar SNe~Ia}
We note that an ejecta mass of $M_{\rm ej,Ch} \approx 1.4$ ($\sim$2 M$_{\odot}$) is likely more appropriate for super-Chandrasekhar explosions, while $E_{\rm K,51} \approx 1$ remains applicable \citep{Scalzo_etal10, Scalzo_etal12, Taubenberger_etal11}. This tweak to the model would serve to make the limits on CSM density less constraining by a factor of $1.2-2.0$ for SN\,2009dc and SN\,2012cu (compared with constraints listed in Tables \ref{snobs} and \ref{snobs_old}).

\subsection{SNe~Ia-CSM}
The strong H$\alpha$ emission lines displayed by SNe~Ia-CSM are evidence of dense CSM, and yet no SN~Ia-CSM has ever been detected at radio wavelengths (out of six observed; Table \ref{snpar}).  Some of this apparent contradiction can be resolved by the fact that SNe~Ia-CSM are rare, and therefore tend to be discovered at large distances; the nearest SN~Ia-CSM in our sample is SN\,2008J at 65 Mpc, and the other four are at 100--300 Mpc. Therefore, the limits on radio luminosity are less deep, compared to more normal SNe~Ia.

\citet{Silverman_etal13} show that the optical light curve rise times of SNe~Ia-CSM are longer than for normal SNe~Ia, and attribute this to interaction with optically thick CSM. From the rise times, they estimate $\dot{M}/v_w \approx$ few $\times10^{-1}~{{{\rm M}_{\odot}\ {\rm yr}^{-1}} \over {{\rm 100\ km\ s}^{-1}}}$ (see also \citealt{Ofek_etal13}). Such dense CSM would lead to the rapid deceleration of the SN ejecta; the models of outer ejecta presented in Section \ref{sec:model} would only apply during the first minutes following explosion, and the SN blast would enter the Sedov-Taylor phase in a matter of years. In addition, substantial free-free absorption would likely dampen the radio signature in the first year, as observed in SNe Type IIn \citep{Chandra_etal12, Chandra_etal15}. A more complete treatment of the evolving blast wave in such dense CSM is required to assess whether radio limits constrain $\dot{M}/v_w \approx$ few $\times10^{-1}~{{{\rm M}_{\odot}\ {\rm yr}^{-1}} \over {{\rm 100\ km\ s}^{-1}}}$, but it is outside the scope of this paper. For now, we simply note that the radio limits on SNe~Ia-CSM presented here are similar to both measurements and upper limits for SNe Type IIn, which are estimated to have $\dot{M}/v_w \approx 10^{-3}-10^{-1}~{{{\rm M}_{\odot}\ {\rm yr}^{-1}} \over {{\rm 100\ km\ s}^{-1}}}$  \citep{vanDyk_etal96, Williams_etal02, Pooley_etal02, Chandra_etal12, Chandra_etal15, Fox_etal13}.

\subsection{iPTF\,14atg}
The SN\,2002es-like event iPTF\,14atg showed evidence for a non-degenerate companion star, based on observations of its early ultraviolet light curve \citep{Cao_etal15}.  \citet{Cao_etal15} estimate that the companion star is located 60--90 R$_{\odot}$ from the explosion site, a separation most consistent with a massive main-sequence or sub-giant companion star. The mass loss rates from such systems would be expected to be significantly lower than for symbiotics, and it is therefore not surprising that iPTF\,14atg went undetected in the radio.

We note that for this explosion, our default assumptions of $M_{\rm ej,Ch} \approx 1$ and $E_{\rm K,51} \approx 1$ are appropriate, given a measured rise time, $t_{\rm rise} = 18.8$ days, expansion velocity of 10,000 km s$^{-1}$ \citep{Cao_etal15}, and Equations 2 and 3 of \citet[][see also \citealt{Arnett82, Pinto_Eastman00a, Pinto_Eastman00b}]{Ganeshalingam_etal12}. Therefore the limits on $\dot{M}/v_w$ and $n_0$ given in Table \ref{snobs_pan} appropriately constrain the CSM in this system.

\subsection{SN\,2006X} \label{sec:06x}

SN\,2006X was the first SN~Ia observed to show time-variable \ion{Na}{1}~D absorption features \citep{Patat_etal07a}, and remains one of a relatively small group in which this phenomenon has been detected ($\sim$3 SNe~Ia; \citealt{Sternberg_etal14}). These absorption features have been interpreted as shells of CSM which are ionized by the SN radiation and then recombine. Estimates of the ionizing flux from SNe~Ia imply the absorbing CSM shells are at radius, $\sim 10^{16}-10^{17}$ cm (\citealt{Patat_etal07a}; see also \citealt{Simon_etal09}). The observed recombination time implies a large electron density in the shells, $\sim 10^{5}$ cm$^{-3}$

Two of the \ion{Na}{1}~D components in SN\,2006X, labeled C and D by \citet{Patat_etal07a}, were observed to increase in depth between 15 days and 31 days after explosion, but had weakened again when next observed 78 days after explosion. The interpretation of these variations offered by Patat et al.\ is that the SN blast plowed over and ionized these components sometime between 31--78 days after explosion. 


This scenario is consistent with our first two epochs of non-detections, observed 6 and 18 days after explosion, as long as the inter-shell material (at radii smaller than shells C and D) is $\lesssim$ few hundred cm$^{-3}$. However, it is outside the scope of the model presented in this paper to asses whether  the third epoch, observed 287--290 days after explosion, is consistent with the SN interaction with a dense shell some $\sim$8 months previous. We look to future work to build off dynamical studies like \citet{Chevalier_Liang89} to predict the radio luminosity and duration from SN--shell interactions.

\subsection{SN\,2011fe}

SN\,2011fe is a remarkably nearby SN~Ia and the subject of an impressive number of multi-wavelength tests to constrain the progenitor system. All came up empty handed (see \citealt{Chomiuk13} for an early review, and further developments by \citealt{Lundqvist_etal15}, \citealt{Graham_etal15c}, and \citealt{Taubenberger_etal15}). More details on the implications of our radio limits for the progenitor system can be read in \citet{Chomiuk_etal12}. 

\subsection{SN\,2012cg}

SN\,2012cg is a core-normal SN~Ia at $\sim$15 Mpc, discovered promptly after explosion \citep{Silverman_etal12, Munari_etal13}. \citet{Marion_etal15} present early-time photometry that shows a blue early-time excess in the optical light curve, consistent with models for SN interaction with a non-degenerate companion \citep{Kasen10}. They show that this excess is best fit with a $>$6 M$_{\odot}$ main sequence or red giant companion. 

After SN\,2011fe and SN\,2014J, our deepest radio limits are for SN\,2012cg, implying $\dot{M}/v_w < 5 \times10^{-9}~{{{\rm M}_{\odot}\ {\rm yr}^{-1}} \over {{\rm 100\ km\ s}^{-1}}}$ for $\epsilon_B = 0.1$, or  $\dot{M}/v_w < 3 \times10^{-8}~{{{\rm M}_{\odot}\ {\rm yr}^{-1}} \over {{\rm 100\ km\ s}^{-1}}}$ for $\epsilon_B = 0.01$. Such low-density CSM is inconsistent with observations of symbiotics (Figure \ref{fig:symbhist}). Most but not all isolated red giants can also be excluded ($\dot{M}$ can extend down to $\sim$ few $\times 10^{-10}$ M$_{\odot}$ yr$^{-1}$ in relatively low-metallicity and unevolved giants; \citealt{Dupree_etal09, Cranmer_Saar11}). Models of main-sequence B stars with $\gtrsim$6 M$_{\odot}$ estimate $\dot{M} =10^{-11}-10^{-9}$ M$_{\odot}$ yr$^{-1}$ and $v_w$ = few thousand km s$^{-1}$ \citep{Krticka14}, which would not be detected by our radio observations. If the mass transfer from the companion to the white dwarf is at all non-conservative (the Kasen models assume the companion is filling its Roche lobe), the CSM around the SN~Ia could be significantly denser than predicted by the companion $\dot{M}/v_w$ alone (see discussion in \citealt{Chomiuk_etal12}). Therefore, our radio observations of SN\,2012cg can constrain but can not completely rule out the non-degenerate companion claimed by \citet{Marion_etal15}.

\subsection{SN\,2014J}

SN\,2014J is the nearest SN~Ia in decades, and our radio non-detections add to an impressive array of strong constraints on the progenitor system \citep{Kelly_etal14, Margutti_etal14, Nielsen_etal14, Perez-Torres_etal14, Lundqvist_etal15}. SN\,2014J is located in a region of high extinction and complex interstellar medium, and several studies do present evidence for material in the vicinity of the system, likely at $\sim 10^{19}$ cm \citep{Foley_etal14, Graham_etal15b, Crotts15}, but it remains unclear if such material is associated with the SN progenitor \citep{Soker15}.

Our radio limits confirm and expand the radio non-detections of \citet{Perez-Torres_etal14}, who utilize the first epoch of VLA non-detections (2014 Jan 23.2) along with additional limits from eMERLIN and EVN obtained 2--5 weeks after outburst. As the earliest epoch is most constraining on a wind CSM profile, we reach very similar limits on $\dot{M}/v_w$ for SN\,2014J. The later observations presented here (3--5 months after explosion) further constrain the presence of CSM at larger radius, $\sim$~several $\times 10^{16}$ cm.

As M82 is a starburst galaxy and SN\,2014J is located at the edge of its inner CO molecular disk \citep{Walter_etal02}, our limits on a uniform density medium around SN\,2014J, $\lesssim 3$ cm$^{-3}$, are plausibly sensitive to the interstellar medium itself. Later time observations will place even deeper constraints on a uniform interstellar medium-like component surrounding SN\,2014J, and can test if the SN exploded in a wind-blown cavity (e.g., \citealt{Badenes_etal07}).

\section{Conclusions}
\begin{itemize}
\item We present observations of 85 thermonuclear SNe observed with the VLA in the first year following explosion; all yield radio non-detections. 

\item This is the most comprehensive study to date of radio emission from thermonuclear SNe. We worked to be complete in collecting VLA observations of thermonuclear SNe, within the parameters described in \S \ref{sec:sample}. These observations are a combination of new data from the Jansky VLA, unpublished archival data, and published limits. 

\item Our models for the radio emission from thermonuclear SNe extend lessons learned from SNe Ib/c to self-consistently predict synchrotron light curves for exploding white dwarf stars. We present models for thermonuclear SN evolution in both wind-stratified and uniform-density material. 

\item We present deep radio limits for SN\,2012cg, with six epochs spanning 5--216 days after explosion, yielding $\dot{M}/v_w \lesssim 5 \times 10^{-9}~{{{\rm M}_{\odot}\ {\rm yr}^{-1}} \over {{\rm 100\ km\ s}^{-1}}}$ for $\epsilon_B = 0.1$. These radio observations are only rivaled by the nearby SN\,2011fe and SN\,2014J.

\item Our sample of thermonuclear SNe spans a range of nine sub-types, including sub-luminous SNe Iax and over-luminous SNe Ia-CSM. In \S \ref{sec:indiv}, we consider appropriate assumptions about ejecta mass and explosion energy for the various sub-types, and modify our model for radio light curves accordingly. 
 
\item The collective radio non-detections imply a scarcity of symbiotic progenitors (i.e., giant companions). We find that $<$10\% of thermonuclear SNe have symbiotic progenitors if we assume $\epsilon_B = 0.1$, or $<$16\% for $\epsilon_B = 0.01$. 
\end{itemize}

In the future, this work can be improved by (a) further observations of Galactic symbiotic binaries, to further pin down their CSM properties; and (b) additional observations of a large number of SNe~Ia with the VLA promptly after explosion, to significantly grow the sample of radio-observed thermonuclear SNe, allowing even stronger constraints on the fraction with red giant companion; and (c) additional radio observations of nearby thermonuclear SNe belonging to sub-types that still have relatively few radio observations (SNe Iax, SN\,2002es-like, super-Chandrasekhar, and Ia-CSM explosions). Further constraints on the CSM and progenitors of SNe Ia can be provided by radio observations at longer times after explosion (1--100 years)---work that our team will present in an upcoming paper.

\acknowledgements 
The National Radio Astronomy Observatory is a facility of the National Science Foundation operated under cooperative agreement by Associated Universities, Inc. We are grateful to J.~McMullin, C.~Chandler, J.~Wrobel, and G.~van Moorsel for their work in commissioning the VLA and scheduling these observations promptly. This work was performed in part at the Aspen Center for Physics, which is supported by National Science Foundation grant PHY-1066293. This research has made use of the NASA/IPAC Extragalactic Database (NED), which is operated by the Jet Propulsion Laboratory, California Institute of Technology, under contract with the National Aeronautics and Space Administration. We are also grateful to David Bishop, who maintains the ``Latest Supernovae" webpage as a member of the Astronomy Section of the Rochester Academy of Science. 

L.~C.\ is a Jansky Fellow of the National Radio Astronomy Observatory. R.~J.~F.\  and A.~M.~S.\ gratefully acknowledge support from the Alfred P.\ Sloan Foundation. J.~S.\ and A.~M.~S.\ are supported by Packard Foundation Fellowships. This research was supported in part by NSF grants AST-0807727 (R.~C.), AST-1412980 (L.~C., S.~B., and C.~B.), and AST-1518052 (R.~J.~F.). 

\bibliography{ref2}

\clearpage
\LongTables
\begin{deluxetable*}{llccccc}
\tabletypesize{\small}
\tablewidth{0 pt}
\tablecaption{ \label{snpar}
 Parameters for Thermonuclear SNe Observed by the VLA}
\tablehead{SN Name & Date of Optical Max & SN Postion & Host Galaxy & Host Type & Distance & SN Ref.\tablenotemark{a} \\
 & & (J2000.0) & & & (Mpc) & }
\startdata

\textbf{Iax:} & \\
SN\,2005hk & 2005 Nov 11 & 00:27:50.87, $-$01:11:52.5 & UGC\,272 & SAB(s)d & 60 & 1 \\
SN\,2008A & 2008 Jan 16 & 01:38:17.38, +35:22:13.7 & NGC\,634 & Sa & 68 & 2 \\
SN\,2008ha & 2008 Nov 13 & 23:34:52.69, +18:13:35.4 & UGC\,12682 & Irr & 20 & 3 \\
SN\,2012Z & 2012 Feb 10 & 03:22:05.35, $-$15:23:15.6 & NGC\,1309 & SA(s)bc & 29 & 4 \\
\\
\hline
\\

\textbf{02es-like:} & \\
iPTF\,14atg\tablenotemark{b} & 2014 May 22 & 12:52:44.8, +26:28:13 & IC 831 & E & 88 & 5\\
\\
\hline
\\

\textbf{Ca-Rich:} & \\
SN\,2000ds & 2000 May 10 & 09:11:36.24, +60:01:42.2 & NGC\,2768 & S0 & 19 & 6\\
SN\,2003H & 2003 Jan 12 & 06:16:25.68, $-$21:22:23.8 & NGC\,2207 & SAB(rs)bc pec & 27 & 7\\
SN\,2003dg & 2003 Apr 10 & 11:57:31.97, $-$01:15:13.6 & UGC\,6934 & SA(r)cd & 84 & 7,8 \\
SN\,2003dr & 2003 Apr 22 & 14:38:11.13, +46:38:03.4 & NGC\,5714 & Sc & 41 & 7,9 \\
SN\,2005E & 2005 Jan 17 & 02:39:14.34, +01:05:55.0 & NGC\,1032 & S0/a & 37 & 10\\
SN\,2005cz & 2005 Jul 2 & 12:37:27.85, +74:11:24.5 & NGC\,4589 & E2 & 29 & 11 \\
SN\,2007ke & 2007 Sep 25 & 02:54:23.90, +41:34:16.3 & NGC\,1129 & E & 73 & 12 \\
PTF\,10iuv\tablenotemark{c} & 2010 Jun 10 & 17:16:54.27, +31:33:51.7 & CGCG\,170-011 & E0 & 98 & 12\\
\\
\hline
\\

\textbf{Cool:} & \\
SN\, 1986G\tablenotemark{d} & 1986 May 11 & 13:25:36.51, $-$43:01:54.2 & NGC\,5128 & S0 pec & 3.8 & 13 \\
SN\,1991bg\tablenotemark{d} & 1991 Dec 13 & 12:25:03.71, +12:52:15.8 & NGC\,4374 & E1 & 17 & 14 \\
SN\, 1999by\tablenotemark{d} & 1999 May 10 & 09:21:52.07, +51:00:06.6 & NGC\,2841 & SA(r)b & 18 & 15\\
SN\,2002cv\tablenotemark{d} & 2002 May 17 & 10 18 03.68 +21 50 06.0 & NGC\,3190 & SA(s)a pec & 25 & 15,16 \\ 
SN\,2003aa & 2003 Feb 11 & 10:46:36.82, +13:45:32.2 & NGC\,3367 & SB(rs)c & 44 & 17 \\
SN\,2005ke & 2005 Nov 23 & 03:35:04.35, $-$24:56:38.8 & NGC\,1371 & SAB(rs)a & 24 & 18 \\
SN\,2006bk & 2006 Apr 1 & 15:04:33.61, +35:57:51.1 & MCG\,+06-33-20 & E & 87 & 19 \\
SN\,2007on\tablenotemark{f} & 2007 Nov 14 & 03:38:50.9, $-$35:34:30 & NGC\,1404 & E1 & 19 & 20 \\
SN\,2011eh & 2011 Jul 31 & 11:18:31.70, +57:58:37.2 & NGC\,3613 & E6 & 29 & 21\\ 
SN\,2011ek\tablenotemark{f} & 2011 Aug 15 & 02:25:48.89, +18:32:00.0 & NGC\,918 & SAB(rs)c & 18 & 22\\
SN\,2011iv\tablenotemark{f} & 2011 Dec 11 & 03:38:51.35, $-$35:35:32.0 & NGC\,1404 & E1 & 19 & 23\\
\\
\hline
\\

\textbf{Broad-line:} & \\
SN\,1980N\tablenotemark{d} & 1980 Dec 12 & 03:22:59.8, $-$37:12:48 & NGC\,1316 & SAB(s)0 pec & 20 & 24\\
SN\,1981B\tablenotemark{d} & 1981 Mar 8 & 12:34:29:57, +02:11:59.3 & NGC\,4536 & SAB(rs)bc & 15 & 25 \\
SN\,1983G\tablenotemark{d} & 1983 Apr 5 & 12:52:20.95, $-$01:12:12.3 & NGC\,4753 & I0 & 18 & 26\\
SN\,1984A\tablenotemark{d} & 1984 Jan 16 & 12:26:55.75, +15:03:18.0 & NGC\,4419 & SB(s)a & 18 & 27 \\
SN\,1987N\tablenotemark{d} & 1987 Dec 19 & 23:19:03.42, $-$08:28:37.7 & NGC\,7606 & SA(s)b & 33 & 28 \\
SN\,1989B\tablenotemark{d} & 1989 Feb 7 & 11:20:13.93, +13:00:19.3 & NGC\,3627 & SAB(s)b & 10 & 29 \\
SN\,1992A\tablenotemark{d} & 1992 Jan 19 & 03:36:27.43, $-$34:57:31.5 & NGC\,1380 & SA0 & 18 & 30 \\
SN\,1995al\tablenotemark{d} & 1995 Nov 8 & 09:50:55.97, +33:33:09.4 & NGC\,3021 & SA(rs)bc & 27 & 31 \\
SN\,1999cl & 1999 Jun 12 & 12:31:56.01, +14:25:35.3 & NGC\,4501 & SA(rs)b & 19 & 32 \\
SN\,1999gh & 1999 Dec 5 & 09:44:19.75, $-$21:16:25.0 & NGC\,2986 & E2 & 33 & 33 \\
SN\,2002bo\tablenotemark{d} & 2002 Mar 24 & 10 18 06.51 +21 49 41.7 & NGC\,3190 & SA(s)a pec & 25 & 34 \\
SN\,2005W & 2005 Feb 11 & 01:50:45.75, +21:45:35.6 & NGC\,691 & SA(rs)bc & 37 & 17 \\
SN\,2006X & 2006 Feb 20 & 12:22:53.99, +15:48:33.1 & NGC\,4321 & SAB(s)bc & 17 & 35 \\
SN\,2006bb & 2006 Apr 1 & 08:33:31.09, +41:31:04.1 & UGC\,4468 & S0 & 96 & 36 \\
SN\,2010ev & 2010 Jul 7 & 10:25:28.99, $-$39:49:51.2 & NGC\,3244 & SA(rs)cd & 30 & 37\\
SN\,2010ko & 2010 Dec 12 & 05:32:49.44, $-$14:05:45.9 & NGC\,1954 & SA(rs)bc & 49 & 38\\
SN\,2011iy & 2011 Dec 3 & 13:08:58.39, $-$15:31:04.1 & NGC\,4984 & (R)SAB(rs)0 & 21 & 39\\
SN\,2012fr & 2012 Nov 12 & 03:33:35.99, $-$36:07:37.7 & NGC\,1365 & SB(s)b & 18 & 40 \\

\\
\hline
\\
\textbf{Core-normal:} & \\
SN\,1985A\tablenotemark{d,g} & 1985 Jan 4 & 09:13:42.40, +76:28:23.9 & NGC\,2748 & SAbc & 21 & 41 \\
SN\,1985B\tablenotemark{d,g} & 1985 Jan 6 & 12:02:44.08, +01:58:45.5 & NGC\,4045 & SAB(r)a & 31 & 41 \\
SN\, 1986A\tablenotemark{d,g} & 1986 Jan 30 & 10:46:36.60, +13:45:01.1 & NGC\,3367 & SB(rs)c & 44 & 42 \\
SN\, 1986O\tablenotemark{d,g} & 1986 Dec 20 & 06:25:58.0, $-$22:00:42 & NGC\,2227 & SB(rs)c & 32 & 43 \\
SN\,1987D\tablenotemark{d} & 1987 Apr 18 & 12:19:41.11, +02:04:26.8 & UGC\,7370 & SBbc pec & 29 & 44 \\
SN\,1989M\tablenotemark{d} & 1989 Jul 2 & 12:37:40.75, +11:49:26.1 & M\,58 & SAB(rs)b & 20 & 45 \\
SN\,1990M\tablenotemark{d} & 1990 Jun 7 & 14:11:29.3,  $-$05:02:36 & NGC\,5493 & S0 pec & 20 & 46 \\
SN\,1994D\tablenotemark{d} & 1994 Mar 20 & 12:34:02.45, +07:42:04.7 & NGC\,4526 & SAB(s)0 & 15 & 47 \\
SN\,1996X\tablenotemark{d} & 1996 Apr 18 & 13:18:01.13, $-$26:50:45.3  & NGC\,5061 & E0 & 25 & 31\\
SN\,1998aq & 1998 Apr 29 & 11:56:26.00, +55:07:38.8 & NGC\,3982 & SAB(r)b & 22 & 48 \\
SN\,1998bu\tablenotemark{d} & 1998 May 19 & 10:46:46.03, +11:50:07.1 & M\,96 & SAB(rs)ab & 11 & 49 \\
SN\,2000E & 2000 Feb 6 & 20:37:13.77, +66:05:50.2 & NGC\,6951 & SAB(rs)bc & 23 & 50 \\
SN\,2001bf & 2001 May 16 & 18:01:33.99, +26:15:02.3 & MCG\,+04-42-22 & ? & 59 & 34\\ 
SN\,2003cg & 2003 Mar 31 & 10:14:15.97, +03:28:02.5 & NGC\,3169 & SA(s)a pec & 26 & 51 \\
SN\,2003du & 2003 May 7 & 14:34:35.80, +59:20:03.8 & UGC\,9391 & SBdm  & 35 & 34 \\
SN\, 2003hv\tablenotemark{d} & 2003 Sep 10 & 03:04:09.32, $-$26:05:07.5 & NGC\,1201 & SA(r)0 & 19 & 34\\
SN\, 2003if\tablenotemark{d,g} & 2003 Sep 1 & 03:19:52.61, $-$26:03:50.5 & NGC\,1302 & (R)SB(r)0/a & 20 & 52\\
SN\,2007sr\tablenotemark{g} & 2007 Dec 14 & 12:01:52.80, $-$18:58:21.7 & NGC\,4038 & SB(s)m pec & 21 & 34 \\
PTF\,10icb & 2010 Jun 14 & 12:54:49.22, +58:52:54.8 & PGC\,43983 & ? & 49 & 53\\
SN\,2010fz & 2010 Jul 21 & 09:42:04.77, +00:19:51.0 & NGC\,2967 & SA(s)c & 31 & 54\\
SN\,2010ih & 2010 Sep 12 & 07:02:43.63, $-$28:37:25.2 & NGC\,2325 & E4 & 22 & 55\\
SN\,2010lq & 2011 Jan 6 & 08:32:40.98, $-$24:02:45.6 & ESO\,495-16 &S0-a &106 & 56\\
SN\,2011B\tablenotemark{g} & 2011 Jan 10 & 08:55:48.50, +78:13:02.7 & NGC\,2655 & SAB(s)0/a & 24 & 57\\
SN\,2011ae & 2011 Feb 27 & 11:54:49.25, $-$16:51:43.6 & PGC\.37373 & SAB(s)cd & 28 & 58\\
SN\,2011at & 2011 Mar 15 & 09:28:57.56, $-$14:48:20.6 & PGC\,26905 & SB(s)d & 25 & 59\\
SN\,2011by & 2011 May 10 & 11:55:45.56, +55:19:33.8 & NGC\,3972 & SA(s)bc & 20 & 21\\
SN\,2011dm & 2011 Jun 27 & 21:56:41.59, +73:17:48.9 & UGC\,11861 & SABdm & 20 & 60\\
SN\,2011fe & 2011 Sep 11 & 14:03:05.81, +54:16:25.4 & M\,101 & SAB(rs)cd & 6.4 & 61\\
SN\,2012cg & 2012 Jun 2 & 12:27:12.83, +09:25:13.2 & NGC\,4424 & SB(s)a & 15 & 62 \\
SN\,2012ei\tablenotemark{g} & 2012 Aug 27 & 14:24:05.71, +33:02:56.5 &NGC\,5611 & S0 & 25 & 63 \\
SN\,2012ht & 2013 Jan 4 & 10:53:22.75, +16:46:34.9 & NGC\,3447 & Pec & 20 & 64 \\
SN\,2013E & 2013 Jan 14 & 10:00:05.52, $-$34:14:01.3 & IC\,2532 & SB(rs)0/a & 32 & 65 \\
SN\,2014J\tablenotemark{g} & 2014 Feb 4 & 09:55:42.14, +69:40:26.0 & M\,82 & Irr & 3.4 & 66 \\
\\
\hline
\\

\textbf{Shallow-Silicon:} & \\
SN\,19991T\tablenotemark{d} & 1991 Apr 26  & 12:34:10.21, +02:39:56.6 & NGC\,4527 & SAB(s)bc & 14 & 67 \\
\\
\hline
\\

\textbf{Super-M$_{\rm ej,Ch}$:} & \\
SN\,2009dc & 2009 Apr 26 & 15:51:12.12, +25:42:28.0 & UGC\,10064 & S0 & 89 & 68 \\
PTF\,10guz & 2010 May 6 & 12:49:22.81, $-$06:23:51.3 & Anonymous & ? & 740 & 69\\
SN\,2012cu & 2012 Jun 27 & 12:53:29.35, +02:09:39.0 &  NGC\,4772 & SA(s)a & 29 & 70 \\
\\
\hline
\\

\textbf{Ia-CSM:} & \\
SN\,1999E & 1999 Jan 1 & 13:17:16.37, $-$18:33:13.4 & GSC\,6116 00964 & Starburst & 104 & 71 \\
SN\,2002ic & 2002 Nov 27 & 01:30:02.55, +21:53:06.9 & NEAT\,J013002.81+215306.9 & ? & 273 & 72\\
SN\,2005gj & 2005 Oct 14 & 03:01:11.95, $-$00:33:13.9 & SDSS\,J030111.99$-$003313.5 & ? & 246 & 72\\
SN\,2008J & 2008 Feb 3 & 02:34:24.20, $-$10:50:38.5 & MCG\,$-$02-7-33 & SBbc & 65 & 72 \\
SN\,2008cg & 2008 Apr 29 & 15:54:15.15, +10:58:25.0 & FGC\,1965 & Scd & 121 & 72 \\
PTF\,11kx\tablenotemark{e} & 2011 Jan 29 & 08:09:12.8, +46:18:48.8	& SDSS J080913.18+461842.9 & Scd & 192 & 73\\
\enddata
\tablenotetext{a}{Light curve maximum estimated from 
1= \citet{Phillips_etal07}; 2= \citet{Foley_etal13}; 3= \citet{Foley_etal09}; 4= \cite{Stritzinger_etal15}; 5=\citet{Cao_etal15}; 6= \citet{00ds}; 7= \citet{03h}; 8= \citet{03dg}; 9= \citet{03dr}; 10= \citet{Perets_etal10}; 11= \citet{Kawabata_etal10}; 12= \citet{Kasliwal_etal12}; 13= \citet{Phillips_etal87}; 14= \citet{Leibundgut_etal93}; 15= \citet{Garnavich_etal04}; 16= \citet{Elias-Rosa_etal08}; 17= \citet{03aa}; 18= \citet{Folatelli_etal10}; 19= \cite{06bk}; 20= \citet{07on}; 21= \cite{Yusa_etal11}; 22= \cite{Maguire_etal12}; 23= \cite{Foley_etal12a}; 24= \citet{80n}; 25= \citet{Schaefer95}; 26= \citet{83g}; 27= \citet{Graham88}; 28= \citet{87n}; 29= \citet{89b}; 30= \citet{92a}; 31= \citet{Riess_etal99}; 32= \citet{Krisciunas_etal06}; 33= \citet{99gh}; 34= \citet{Ganeshalingam_etal10}; 35= \citet{Wang_etal08}; 36= \citet{06bb}; 37= \citet{10ev}; 38= \cite{Challis_etal10}; 39= \cite{Noguchi_etal11}; 40= \cite{Childress_etal13}; 41= \citet{Wegner87}; 42= \citet{86a}; 43= \citet{Arsenault88}; 44= \citet{87d}; 45= \citet{89m}; 46= \citet{dellaValle_etal96};  47= \citet{94d}; 48= \citet{98aq};  49= \citet{98bu}; 50= \citet{00e}; 51= \citet{03cg}; 52= \citet{03if}; 53= \cite{Nugent_etal10_10icb}; 54= \cite{Prieto_etal10}; 55= \cite{Morrell_etal10}; 56= \cite{Prieto_Morrell11}; 57= \cite{Yamaoka_etal11}; 58= \cite{11ae}; 59= \cite{11at}; 60= \cite{Kandrashoff_etal11}; 61= \cite{11fe};  62= \citet{Silverman_etal12}; 63= \cite{Nakano_etal12}; 64= \cite{Yamanaka_etal14}; 65= \cite{Kiyota_etal13}; 66= \cite{Marion_etal15}; 67= \cite{91t};  68= \citet{Taubenberger_etal11}; 69= \cite{Nugent_etal10_10guz}; 70= \cite{Amanullah_etal15}; 71= \citet{99e}, discovered on 1999 Jan 15 after maximum. Time of optical maximum is very uncertain; 72= \citet{Silverman_etal13};  73=\citet{Dilday_etal12}.}
\tablenotetext{}{Radio observations published in $^{b}$\cite{Cao_etal15}; $^{c}$\cite{Kasliwal_etal12}; $^{d}$\cite{Panagia_etal06}; $^{e}$\cite{Dilday_etal12}.}
\tablenotetext{f}{Hybrid cool/normal.}
\tablenotetext{g}{There is some ambiguity for this SN between core-normal and broad-line classifications.}
\end{deluxetable*} 


\clearpage
\begin{deluxetable}{llcccccccc}
\tabletypesize{\small}
\tablewidth{0 pt}
\tablecaption{ \label{snobs}
Observations of Thermonuclear SNe with the Upgraded VLA}
\tablehead{SN Name & Observation & VLA & Central & Band- & Flux & Time Since & Luminosity & $\dot{M}/v_{w}$\tablenotemark{a} & $n_0$\tablenotemark{a}\\
 & Date & Config. & Freq. & width & Density & Explosion &  Upper Limit & Upper Limit & Upper Limit \\
 & (UT) & & (GHz) & (MHz) & ($\mu$Jy) & (Days) & ($10^{25}$ erg s$^{-1}$ Hz$^{-1}$) &  $\left({{{\rm M}_{\odot}\, {\rm yr}^{-1}} \over {100\ {\rm km\, s}^{-1}}}\right)$ & (cm$^{-3}$) }
\startdata
PTF\,10guz\tablenotemark{b} & 2010 May 11.1 & D & 4.9 & 256 & $-16 \pm 23$  & 29.1 & 4521.83 & N/A & N/A \\

PTF\,10icb & 2010 Jun 4.0 & D & 4.9 & 256 & $164 \pm 71$ & 8.0 & 108.33 & N/A & N/A \\
\nodata & 2010 Jun 6.0 & D & 7.9 & 128 & $7 \pm 27$ & 10.0 & 25.29 & $3.2 \times 10^{-7}$ & 2500\\
\nodata & 2010 Jun 8.1 & D & 8.4 & 256 & $35 \pm 22$ & 12.1 & 29.02 & $4.0 \times 10^{-7}$ & 2500 \\
\nodata & 2010 Jun 11.1 & D & 7.9 & 128 & $32 \pm 26$ & 15.1 & 31.60 & $5.0 \times 10^{-7}$ & 2500 \\

SN\,2010ev & 2010 Jul 3.9 & D & 5.0 & 1024 & $29 \pm 30$ & 13.5 & 12.82 & $1.6 \times 10^{-7}$ & 1000 \\

SN\,2010fz\tablenotemark{c} & 2010 Jul 11.9 & D & 6.0 & 2048 & $-13 \pm 9$ & 8.9 & 3.11 & $5.0 \times 10^{-8}$ & 400 \\

SN\,2010ih & 2010 Oct 15.5 & D & 5.9 & 2048 & $-18 \pm 13$ & 51.5 & 2.26 & $2.0 \times 10^{-7}$ & 100 \\

SN\,2010ko & 2010 Dec 9.3 & C & 5.9 & 2048 & $-11 \pm 6$ & 13.9 & 5.17 & $1.0 \times 10^{-7}$ & 500 \\

SN\,2010lq & 2011 Jan 21.3 & CnB & 4.9 & 256 & $7 \pm 17$ & 33.3 & 77.99 & $1.3 \times 10^{-6}$ & 2000 \\

SN\,2011B & 2011 Jan 21.1 & CnB & 5.9 & 2048 & $2 \pm 9$ & 29.1 & 1.99 & $1.3 \times 10^{-7}$ & 130 \\

SN\,2011ae & 2011 Mar 11.3 & B & 5.9 & 2048 & $-5 \pm 5$ & 30.3 & 1.41 & $1.0 \times 10^{-7}$ & 100\\

SN\,2011at & 2011 Mar 17.2 & B & 5.9 & 2048 & $-1 \pm 5$  & 20.2 & 1.12 & $5.0 \times 10^{-8}$ & 100 \\

SN\,2011by & 2011 Apr 28.0 & B  & 5.9 & 2048 & $16 \pm 6$ & 6.0 & 1.63 & $2.0 \times 10^{-8}$ & 250 \\
\nodata & 2011 Apr 30.1 & B & 5.9 & 2048 & $1 \pm 3$ & 8.1 & 0.48 & $1.0 \times 10^{-8}$ & 63 \\
\nodata & 2011 Jun 10.1 & A & 5.9 & 2048  & $11 \pm 7$  & 49.1 & 1.53 & $1.6 \times 10^{-7}$ & 79 \\

SN\,2011dm & 2011 Jun 23.1 & A & 5.9 & 2048 & $-4 \pm 5$  & 14.1 & 0.72 & $2.5 \times 10^{-8}$ & 79\\

SN\,2011eh & 2011 Aug 4.5 & A & 5.9 &  2048 & $-3 \pm 7$  & 17.5 & 2.11 & $7.9 \times 10^{-8}$ & 200 \\

SN\,2011ek & 2011 Aug 9.4 & A & 5.9 &  2048 & $0 \pm 5$  & 7.4 & 0.58 &$2.0 \times 10^{-8}$  & 79 \\

SN\,2011fe\tablenotemark{d} & 2011 Aug 25.8 & A & 5.9 &  2048 &$8\pm6$ & 1.8 & 0.12 & $7.9 \times 10^{-10}$ & 32 \\
\nodata\tablenotemark{d} & 2011 Aug 27.7 & A & 5.9 &  2048 &$0\pm6$ & 3.7 & 0.09 & $1.3 \times 10^{-9}$ & 20 \\
\nodata\tablenotemark{d} & 2011 Aug 30.0 & A & 5.9 &  2048 &$-8\pm7$ & 6.0 & 0.10 & $2.5 \times 10^{-9}$ & 20 \\
\nodata\tablenotemark{d} & 2011 Sept 2.7 & A & 5.9 &  2048 &$1\pm4$ & 9.7 & 0.07 & $3.2 \times 10^{-9}$ & 10 \\
\nodata\tablenotemark{d} & 2011 Sept 8.0 & A & 5.9 &  2048 &$-3\pm7$ & 15.0 & 0.10 & $6.3 \times 10^{-9}$ & 13\\
\nodata\tablenotemark{d} & 2011 Sept 12.9 & A & 5.9 &  2048 &$1 \pm6$ & 19.9 & 0.09 & $1.0 \times 10^{-8}$ & 13 \\

SN\,2011iv & 2011 Dec 4.2 & D & 6.8 & 1024 & $18 \pm 18$  & 6.2 & 3.11 & $3.2 \times 10^{-8}$ & 500\\

SN\,2011iy & 2011 Dec 20.5 & C & 7.3 & 1024 & $-39\pm45$ & 34.1 & 7.12 & $4.0 \times 10^{-7}$ & 400\\

SN\,2012Z & 2012 Feb 2.0 & C & 5.9 & 2048 & $-15\pm6$ & 7.0 & 1.81 & $2.5 \times 10^{-8}$ & 250\\

SN\,2012cg\tablenotemark{e} & 2012 May 19.1 & CnB & 4.1 & 2048 & $-14\pm5$ & 5 & 0.40 & $5.0 \times 10^{-9}$ & 63\\
\nodata\tablenotemark{e} & 2012 Jun 27.1 & B & 5.9 & 2048 & $-5\pm5$ & 43.1 & 0.40 & $6.3 \times 10^{-8}$ & 32\\
\nodata\tablenotemark{e} & 2012 Jul 19.0 & B & 5.9 & 2048 & $-12\pm7$ & 65.0 & 0.58 & $1.3 \times 10^{-7}$ & 32\\
\nodata\tablenotemark{e} & 2012 Aug 26.0 & B & 5.9 & 2048 & $-16\pm6$ & 103.0 & 0.48 & $1.6 \times 10^{-7}$ & 20\\
\nodata & 2012 Oct 21.8 & A & 5.9 & 2048 & $-1\pm5$ & 159.8 & 0.40 & $2.0 \times 10^{-7}$ & 13\\
\nodata & 2012 Dec 16.6 & A & 5.9 & 2048 & $-13\pm5$ & 215.6 & 0.40 & $2.5 \times 10^{-7}$ & 10 \\

SN\,2012cu & 2012 Jun 18.2 & B & 5.9 & 2048 & $1\pm5$ & 15.2 & 1.61 & $5.0 \times 10^{-8}$ & 160 \\

SN\,2012ei & 2012 Aug 25.0 & B & 5.9 & 2048 & $-2\pm6$ & 16.0 & 1.35 & $5.0 \times 10^{-8}$ & 130\\

SN\,2012fr & 2012 Oct 30.3 & A & 5.9 & 2048 & $8\pm7$ & 3.9 & 1.12 & $7.9 \times 10^{-9}$ & 200\\

SN\,2012ht & 2012 Dec 21.6 & A & 5.9 & 2048 & $4\pm5$ & 4.6 & 0.91 & $7.9 \times 10^{-9}$ & 160\\
\nodata & 2013 Jan 4.5 & A & 5.9 & 2048 & $0\pm5$ & 18.5 & 0.72 & $3.2 \times 10^{-8}$ & 80\\

SN\,2013E & 2013 Jan 8.4 & A$\rightarrow$D & 5.9 & 2048 & $-7\pm8$ & 12.4 & 2.94 & $6.3 \times 10^{-8}$ & 320\\

SN\,2014J\tablenotemark{f} & 2014 Jan 23.2 & B$\rightarrow$BnA & 5.5 & 2048 & $-12\pm8$ & 6.2 & 0.03 & $1.0 \times 10^{-9}$ & 6.3 \\
\nodata & 2014 Apr 11.2 & A & 5.9 & 2048 & $0\pm7$ & 84.2 & 0.03 & $2.0 \times 10^{-8}$ & 2.5 \\
\nodata & 2014 Jun 12.0 & A$\rightarrow$D& 7.4 & 1024 & $11\pm10$ & 146.0 & 0.06 & $6.3 \times 10^{-8}$ & 3.2 \\

\enddata
\tablenotetext{a}{Assuming $\epsilon_B = 0.1$, $M_{\rm ej,Ch}=1$, and $E_{\rm K,51}=1$.}
\tablenotetext{}{Radio limit originally published in $^{\rm b}$\cite{Chomiuk10guz}; $^{\rm c}$\cite{Chomiuk10fz}; $^{\rm d}$\cite{Chomiuk_etal12} (see also \citealt{Horesh_etal11}); $^{\rm e}$\cite{Chomiuk12cg}; $^{\rm f}$\cite{Chomiuk14j} (see also \citealt{Chandler_Marvil14} and \citealt{Perez-Torres_etal14}).}
\end{deluxetable}

\clearpage
\LongTables
\begin{deluxetable}{llcccccccc}
\tabletypesize{\small}
\tablewidth{0 pt}
\tablecaption{ \label{snobs_old}
Archival Observations of Thermonuclear SNe with the Pre-Upgrade VLA}
\tablehead{SN Name & Observation & VLA & Central & Program & Flux & Time Since & Luminosity & $\dot{M}$/v$_{w}$\tablenotemark{a} & $n_0$\tablenotemark{a} \\
 & Date & Config. & Freq. & ID & Density & Explosion &  Upper Limit & Upper Limit & Upper Limit \\
 & (UT) & & (GHz) & & ($\mu$Jy) & (Days) & ($10^{25}$ erg s$^{-1}$ Hz$^{-1}$) &  $\left({{{\rm M}_{\odot}\, {\rm yr}^{-1}} \over {100\ {\rm km\, s}^{-1}}}\right)$ & (cm$^{-3}$)}
\startdata

SN\,1998aq & 1998 Jun 27.7 & BnA & 4.9 & AL446 & $56\pm 71$ & 77.7 & 15.58 & $1.0 \times 10^{-6}$ & 250\\

SN\,1999E & 1999 Jan 22.3 & C & 8.4 & ACTST & $145\pm 80$ & 51.3 & 498.34 & $7.9 \times 10^{-6}$ & 10,000\\
\nodata & 1999 Jan 22.3 & C & 14.9 & ACTST & $439\pm 319$ & 51.3 & 1806.98 & $3.2 \times 10^{-5}$ & 40,000\tablenotemark{k} \\
\nodata & 1999 Feb 8.4 & DnC & 8.4  & AS568 & $50\pm 46$ & 68.4 & 243.35 & $6.3 \times 10^{-6}$ & 4000\\
\nodata & 1999 Feb 8.4 & DnC & 14.9 & AS568 & $140\pm 172$ & 68.4 & 849.13 & $2.0 \times 10^{-5}$ & 16,000\tablenotemark{k} \\
\nodata & 1999 Feb 8.4 & DnC & 22.4 & AS568 & $-295\pm 281$ & 68.4 & 1091.18 & $2.5 \times 10^{-5}$ & 25,000\tablenotemark{k} \\
\nodata & 1999 Feb 27.3 & DnC & 8.4 & AS568 & $23\pm 50$ & 87.3 & 223.93 & $7.9 \times 10^{-6}$ & 3200 \\
\nodata & 1999 Feb 27.3 & DnC & 14.9 & AS568 & $-226\pm 169$ & 87.3 & 656.26 & $2.0 \times 10^{-5}$ & 10,000\tablenotemark{k} \\
\nodata & 1999 Feb 27.3 & DnC & 22.4 & AS568 & $-169\pm 175$ & 87.3 & 679.56 & $2.5 \times 10^{-5}$ & 16,000\tablenotemark{k} \\
\nodata & 1999 Oct 28.8 & BnA & 8.4 & AS568 & $-36\pm 47$ & 330.8 & 182.51 & $2.0 \times 10^{-5}$ \tablenotemark{k} & 1000\tablenotemark{k}  \\

SN\,1999cl & 1999 Oct 31.7 & B & 4.9 & AU79 & $-39\pm 61$ & 158.3 & 7.91 & $1.0 \times 10^{-6}$ & 100\\

SN\,1999gh & 2000 Mar 14.1 & BnC & 8.4 & AJ275 & $35\pm 65$ & 115.7 & 29.97 & $2.5 \times 10^{-6}$ & 500\\

SN\,2000E & 2000 Oct 8.0 & D & 4.9 & AS568 & $-21\pm 60$ & 263.0 & 11.40 & $2.0 \times 10^{-6}$ & 100\\
\nodata & 2000 Oct 17.2 & A & 4.9 & AS568 & $-79\pm 44$ & 272.2 & 8.36 & $1.6 \times 10^{-6}$ & 63\\

SN 2000ds & 2000 Oct 29.3 & A & 8.4 & AK509 & $42\pm 54$ & 184.3 & 8.81 & $2.0 \times 10^{-6}$ & 160 \\
\nodata & 2001 Feb 19.4  & BnA & 8.4 & AF380 & $-56\pm 24$ & 297.4 & 3.11 &  $1.6 \times 10^{-6}$ & 50\\

SN\,2001bf & 2001 Jun 1.3 & ? & 8.4 & AK509 & $-21\pm 49$ & 34.3 & 61.24 & $1.6 \times 10^{-6}$ & 2000\\

SN\,2002ic\tablenotemark{b} & 2003 Jun 16.6 & A & 4.9 & AK550 & $25\pm 33$ & 231.6 & 1105.98 & $3.2 \times 10^{-5}$ \tablenotemark{k}  & 3200\tablenotemark{k} \\
\nodata \tablenotemark{c} & 2003 Jun 17.8 & A & 22.4 & AW608 & $-202\pm 157$ & 232.8 & 4200.95 & $1.5 \times 10^{-4}$  \tablenotemark{k}& 32,000\tablenotemark{k} \\

SN 2003H & 2003 Jan 14.3 & DnC & 8.4 & AK509 & $-27\pm 40$ & 14.3 & 10.47 & $2.5 \times 10^{-7}$ & 1000\\

SN\,2003aa & 2003 Feb 3.3 & DnC & 8.4 & AK550 & $29\pm 43$ & 5.3 & 36.61 & $3.2 \times 10^{-7}$ & 7900 \\
\nodata & 2003 Jun 15.0 & A & 8.4 & AK509 & $36\pm 69$ & 137.0 & 56.30 & $5.0 \times 10^{-6}$ & 790\\

SN\,2003cg & 2003 May 17--29 & A & 8.4 & AA278 & $88\pm 38$ & 65--77 & 16.34 & $1.3 \times 10^{-6}$ & 500\\
\nodata & 2003 Sep 12--18 & A & 8.4 & AA278 & $-70\pm 65$ & 183--189 & 15.78 &  $2.5 \times 10^{-6}$ & 250\\
\nodata & 2003 Sep 21--25 & BnA & 8.4 & AA278 & $8\pm 50$ & 192--196 & 12.78 & $2.5 \times 10^{-6}$ & 200\\

SN\,2003dg & 2003 July 1.0 & A & 4.9 & AK509 & $-8\pm 66$ & 94.0 & 128.35 & $5.0 \times 10^{-6}$ & 2000\\
\nodata & 2003 July 1.0 & A & 8.4 & AK509 & $8\pm 48$ & 94.0 & 167.20 & $5.0 \times 10^{-6}$ & 1600 \\
\nodata & 2003 Nov 1.7 & B & 8.4 & AK573 & $154\pm 77$ & 217.7 & 325.10 & $2.0 \times 10^{-5}$ \tablenotemark{k} & 2000\tablenotemark{k}  \\

SN\,2003dr & 2003 July 1.0 & A & 4.9 & AK509 & $81\pm 68$ & 82.0 & 28.37 & $2.0 \times 10^{-6}$ & 630\\
\nodata& 2003 July 1.0 & A & 8.4 & AK509 & $-111\pm 47$ & 82.0 & 57.33 & $2.0 \times 10^{-6}$ & 790\\
\nodata & 2003 Nov 1.7 & B & 8.4 & AK573 & $28\pm 65$ & 205.7 & 44.86 & $5.0 \times 10^{-6}$ & 500\tablenotemark{k} \\

SN\,2003du & 2003 Oct 17.0 & BnA & 8.4 & AS779 & $-76\pm 33$ & 181.0 & 14.51 & $2.5 \times 10^{-6}$ & 200\\

SN 2005E\tablenotemark{d} & 2005 Jan 21.1 & BnA & 8.4 & AS796 & $11\pm 53$ & 16.1 & 27.85 & $5.0 \times 10^{-7}$ & 2000\\

SN\,2005W & 2005 Feb 7.9 & BnA & 8.4 & AW641 & $-50\pm 52$ & 13.5 & 25.56 & $4.0 \times 10^{-7}$ & 2500 \\

SN 2005cz & 2005 Aug 7.6 & C & 8.4 & AS846 & $110\pm 65$ & 48.6 & 30.70 & $1.3 \times 10^{-6}$ & 1000 \\
\nodata & 2006 Apr 24.1 & A & 4.9 & AC807 & $16\pm 61$ & 308.1 & 20.03 & $3.2 \times 10^{-6}$ & 130\\

SN\,2005gj\tablenotemark{e} & 2005 Nov 25.4 & D & 8.4 & AS869 & $26\pm 32$ & 72.4 & 883.55 & $1.6 \times 10^{-5}$ & 10,000 \\
\nodata & 2005 Dec 6.1 & D & 8.4 & AK583 & $75\pm 40$ & 83.1 & 1412.23 & $2.5 \times 10^{-5}$ & 16,000\tablenotemark{k} \\
\nodata & 2006 Jan 2.1 & D & 8.4 & AK583 & $45\pm 25$ & 110.1& 869.07 & $2.0 \times 10^{-5}$ & 7900\tablenotemark{k} \\
\nodata & 2006 Feb 26.9 & A & 1.4 & AW675 & $-29\pm 52$ & 165.9 & 1129.79 &  N/A & N/A\\
\nodata & 2006 Feb 26.9 & A & 8.4 & AW675 & $-12\pm 35$ & 165.9 & 760.43 & $2.5 \times 10^{-5}$ \tablenotemark{k} & 5000\tablenotemark{k}  \\

SN\,2005hk & 2006 Jan 31.0 & A & 8.4 & AS846 & $77\pm 29$ & 96.0 & 70.66 & $4.0 \times 10^{-6}$ & 1300\\
\nodata & 2006 Feb 4.0 & A & 4.9 & AS846 & $19\pm 96$  & 100.0 & 132.26 & $4.0 \times 10^{-6}$ & 1300\\
\nodata & 2006 Feb 4.0 & A & 8.4 & AS846 & $131\pm 90$  & 100.0 & 172.76 & $6.3 \times 10^{-6}$ & 2000\\

SN\,2005ke\tablenotemark{f} & 2005 Nov 18.2 & D & 8.4 & AS846 & $70\pm 59$ & 8.2 & 17.03 & $2.0 \times 10^{-7}$ & 2500\\
\nodata\tablenotemark{f} & 2006 Jan 27.1 & A & 8.4 & AS846 & $21\pm 58$ & 78.1 & 13.44 & $1.3 \times 10^{-6}$ & 400\\

SN\,2006X & 2006 Feb 9.3 & A & 8.4 & AS875 & $-27\pm 18$ & 5.9 & 1.87 & $2.5 \times 10^{-8}$ & 400\\
\nodata & 2006 Feb 9.4 & A & 22.4 & AW682 & $75\pm 58$ & 6.0 & 8.61 & $1.6 \times 10^{-7}$ & 3200\\
\nodata & 2006 Feb 21.4 & A & 1.4 & AW682 & $-14\pm 54$ & 18.0 & 5.60 & $6.3 \times 10^{-8}$ & 160 \\
\nodata & 2006 Feb 21.4 & A & 8.4 & AW682 & $-26\pm 24$ & 18.0 & 2.49 & $1.0 \times 10^{-7}$ & 320 \\
\nodata\tablenotemark{g} & 2006 Nov 17.6 & C & 4.8 & AC867 & $-36\pm 42$ & 287.2 & 4.36 & $1.3 \times 10^{-6}$ & 40\\
\nodata\tablenotemark{g} & 2006 Nov 20.6 & C & 8.4 & AC867 & $-19\pm 36$ & 290.2 & 3.74 & $1.6 \times 10^{-6}$ & 50\\

SN\,2006bb & 2006 Apr 3.1 & A & 8.4 & AK583 & $79\pm 58$ & 18.7 & 279.04 & N/A & N/A\\

SN\,2006bk & 2006 Apr 13.4 & A & 8.4 & AS877 & $5\pm 48$ & 25.4 & 134.97 & $2.0 \times 10^{-6}$ & 5000\tablenotemark{k}  \\

SN 2007ke & 2007 Oct 24.4 & BnA & 8.4 & AS887 & $-32\pm 36$ & 41.4 & 68.88 & $2.0 \times 10^{-6}$ & 2000\\

SN\,2007on\tablenotemark{h} &  2007 Dec 7.3 & B & 1.4 & AS935 & $-95\pm 535$ & 36.3 & 69.34 & N/A & N/A \\
\nodata\tablenotemark{h} &  2007 Dec 7.2 & B & 8.4 & AS935 & $3\pm 33$ & 36.2 & 4.41 & $3.2 \times 10^{-7}$ & 250\\

SN\,2007sr & 2007 Dec 23.4 & B & 8.4 & AS887 & $33\pm 48$ & 27.4 & 9.34 & $4.0 \times 10^{-7}$ & 630\\

SN\,2008A & 2008 Jan 6.1 & B & 8.4 & AS887 & $-59\pm 46$ & 5.1 & 76.37 & N/A & N/A \\

SN\,2008J & 2008 Mar 1.8 & CnB & 8.4 & AS887 & $-36\pm 49$ & 57.8 & 74.33 &  $2.5 \times 10^{-6}$ & 1600 \\
\nodata & 2008 Sep 25.4 & D & 8.4 & AC938 & $119\pm 107$ & 265.4 & 222.47 & $2.0 \times 10^{-5}$ \tablenotemark{k} & 1300\tablenotemark{k}  \\
\nodata & 2008 Dec 16.2 & A & 8.4 & AC938 & $96\pm 62$ & 347.2 & 142.59 & $1.6 \times 10^{-5}$ \tablenotemark{k} & 790\tablenotemark{k}  \\

SN\,2008cg & 2008 Jun 13.1 & DnC & 8.4 & AC881 & $30\pm 55$ & 75.1 & 341.67 & $7.9 \times 10^{-6}$ & 5000\\
\nodata\tablenotemark{i} & 2008 Jul 1.2 & D & 8.4 & AC881 & $7\pm 38$ & 93.2 & 212.01 & $7.9 \times 10^{-6}$ & 2500\\
\nodata & 2008 Dec 13.6 & A & 8.4 & AC938 & $60\pm 52$ & 258.6 & 378.47 & $2.5 \times 10^{-5}$ \tablenotemark{k} & 2000\tablenotemark{k} \\

SN\,2008ha\tablenotemark{j} & 2008 Nov 22.0 & A & 8.4 & AS929 & $-23\pm 48$ & 24.0 & 6.89 & $3.2 \times 10^{-7}$ & 500 \\

SN\,2009dc & 2009 Apr 28.5 & B & 8.4 & AS929 & $24\pm 48$ & 26.5 & 159.25 & $2.5 \times 10^{-6}$ & 6300\\
\enddata
\tablenotetext{a}{Assuming $\epsilon_B = 0.1$, $M_{\rm ej,Ch}=1$, and $E_{\rm K,51}=1$.}
\tablenotetext{}{Radio observations originally published in $^{b}$\citet{02ic_radio1}; $^{c}$\citet{02ic_radio2}; $^{\rm d}$\citet{Perets_etal10}; $^{\rm e}$\citet{Soderberg05gj}; $^{\rm f}$\citet{Soderberg05ke}; $^{\rm g}$\citet{Chandra06x}; $^{\rm h}$\citet{Soderberg07on}; $^{\rm i}$\citet{Chandra08cg}; $^{j}$\citet{Soderberg08ha}.}
\tablenotetext{k}{This limit is not sufficiently constraining to be completely described by the model of outer SN ejecta applied in this study. This radio limit is consistent with dense CSM which leads to rapid deceleration of the SN blast. It is therefore possible that the steep outer ejecta are not interacting with the CSM at the time of this observation, but rather the inner shallower ejecta are interacting.}
\end{deluxetable}

\LongTables
\begin{deluxetable}{llccccccc}
\tabletypesize{\small}
\tablewidth{0 pt}
\tablecaption{ \label{snobs_pan}
Published Observations of SNe~Ia with the VLA}
\tablehead{SN Name & Observation & VLA & Central  & $\sigma_{\nu}$ & Time Since & Luminosity & $\dot{M}$/v$_{w}$\tablenotemark{a} & $n_0$\tablenotemark{a}\\
 & Date & Config. & Freq. &  & Explosion &  Upper Limit & Upper Limit & Upper Limit \\
 & (UT) & & (GHz) & (mJy) & (Days) & ($10^{25}$ erg s$^{-1}$ Hz$^{-1}$) &  $\left({{{\rm M}_{\odot}\, {\rm yr}^{-1}} \over {100\ {\rm km\, s}^{-1}}}\right)$ & (cm$^{-3}$)}
\startdata
SN\,1980N & 1981 Feb 3 & A & 4.8 & 0.20 & 70 & 23.37 & $1.0 \times 10^{-6}$ & 400\\   		
SN\,1981B & 1981 Mar 11 & A & 4.8 & 0.10 & 20 & 6.57 & $1.6 \times 10^{-7}$ & 400\\   
\nodata & 1981 Apr 9 & A & 4.8 & 0.13 & 49 & 8.54 & $4.0 \times 10^{-7}$ & 250\\   
\nodata & 1981 May 14 & B & 4.8 & 0.17 & 84 & 11.17 & $7.9 \times 10^{-7}$ & 200 \\   
\nodata & 1981 Jun 19 & B & 4.8 & 0.20 & 120 & 13.14 & $1.3 \times 10^{-6}$ & 200\\   
\nodata & 1981 Aug 13 & B & 4.8 & 0.30 & 175 & 19.72 & $2.0 \times 10^{-6}$ & 200\\   
\nodata & 1981 Nov 11 & C & 4.8 & 0.07 & 265 & 4.60 & $1.3 \times 10^{-6}$ & 50\\   
SN\,1983G & 1983 May 27 & C & 4.8 & 0.07 & 69 & 6.62 & $5.0 \times 10^{-7}$ & 160\\  
SN\,1984A & 1984 Mar 5 & BnC & 4.8 & 0.10 & 66 & 9.46 & $6.3 \times 10^{-7}$ & 200\\  
SN\,1985A & 1985 Feb 1 & A & 4.8 & 0.07 & 46 & 9.02 & $4.0 \times 10^{-7}$ & 250\\ 
\nodata & 1985 Feb 8 & A & 1.4 & 0.07 & 53 & 2.63 & $1.0 \times 10^{-7}$ & 40\\ 
\nodata & 1985 Feb 17 & A & 1.4 & 0.09 & 62 & 3.38 & $1.3 \times 10^{-7}$ & 40\\ 
\nodata & 1985 Feb 17 & A & 4.8 & 0.07 & 62 & 9.02 & $5.0 \times 10^{-7}$ & 200\\ 
\nodata & 1985 Mar 2 & A & 1.4 & 0.07 & 75 & 2.63 & $1.3 \times 10^{-7}$ & 32\\ 
\nodata & 1985 Apr 5 & A/B & 1.4 & 0.08 & 109 & 3.00 & $2.0 \times 10^{-7}$ & 25\\ 
\nodata & 1985 Oct 28 & C/D & 1.4 & 0.10 & 315 & 3.76 & $6.3 \times 10^{-7}$ & 13\\ 
SN\,1985B & 1985 Feb 22 & A & 1.4 & 0.17 & 65 & 13.92 & $4.0 \times 10^{-7}$ & 130 \\ 
\nodata & 1985 Feb 22 & A & 4.8 & 0.19 & 65 & 53.33 & $1.6 \times 10^{-6}$ & 790\\ 
\nodata & 1985 Mar 18 & A/B & 1.4 & 0.40 & 89 & 32.75 & $7.9 \times 10^{-7}$ & 200\\ 
\nodata & 1985 Mar 18 & A/B & 4.8 & 0.06 & 89 & 16.84 & $1.0 \times 10^{-6}$ & 250\\ 
\nodata & 1985 Sep 15 & C & 1.4 & 0.08 & 270 & 6.55 & $7.9 \times 10^{-7}$ & 25\\ 
\nodata & 1985 Sep 15 & C & 4.8 & 0.05 & 270 & 14.03 & $2.5 \times 10^{-6}$ & 100\\ 
SN\,1986A & 1986 Feb 7 & D & 4.8 & 0.22 & 26 & 124.41 & $2.0 \times 10^{-6}$ & 4000\\ 
\nodata & 1986 Feb 25 & A & 1.4 & 0.36 & 44 & 59.38 & N/A & N/A\\ 
\nodata & 1986 Feb 25 & A & 4.8 & 0.12 & 44 & 67.86 & $1.6 \times 10^{-6}$ & 1300\\ 
\nodata & 1986 Mar 16 & A & 4.8 & 0.05 & 63 & 28.27 & $1.0 \times 10^{-6}$ & 500\\ 
\nodata & 1986 Apr 3 & A & 4.8 & 0.09 & 81 & 50.89 & $2.0 \times 10^{-6}$ & 630\\ 
\nodata & 1986 Jun 15 & A/B & 1.4 & 0.15 & 154 & 24.74 & $1.0 \times 10^{-6}$ & 100\\ 
\nodata & 1986 Jun 15 & A/B & 4.8 & 0.09 & 154 & 50.89 & $3.2 \times 10^{-6}$ & 400\\ 
\nodata & 1986 Oct 16 & B/C & 1.4 & 0.25 & 277 & 41.23 & $2.5 \times 10^{-6}$ & 100\\ 
\nodata & 1986 Oct 16 & B/C & 4.8 & 0.08 & 277 & 45.24 & $5.0 \times 10^{-6}$ & 250\tablenotemark{g}\\ 
SN\,1986G & 1986 May 14 & A & 4.8 & 1.06 & 16 & 4.47 & $1.0 \times 10^{-7}$ & 320\\    
\nodata & 1986 May 21 & A & 4.8 & 0.70 & 23 & 2.95 & $1.0 \times 10^{-7}$ & 200\\
\nodata & 1986 May 21 & A & 15.0 & 3.21 & 23 & 42.31 &$1.3 \times 10^{-6}$ & 3200\\
\nodata & 1986 Jun 8 & A & 4.8 & 1.02 & 41 & 4.30 & $2.5 \times 10^{-7}$ & 160\\
\nodata & 1986 Jul 6 & A/B & 4.8 & 1.52 & 69 & 6.41& $5.0 \times 10^{-7}$ & 160\\
\nodata & 1986 Jul 6 & A/B & 15.0 & 1.32 & 69 & 17.40 &$1.6 \times 10^{-6}$ & 790 \\
\nodata & 1986 Sep 20 & C & 4.8 & 2.49 & 145 & 10.50& $1.3 \times 10^{-6}$ & 130 \\
\nodata & 1986 Sep 20 & C & 15.0 & 1.27 & 145 & 16.74 &$3.2 \times 10^{-6}$ & 400 \\
\nodata & 1987 Jan 4 & C & 4.8 & 2.71 & 251 & 11.43 & $2.0 \times 10^{-6}$ & 100 \\
\nodata & 1987 Jan 4 & C & 15.0 & 1.38 & 251 & 18.19 & $5.0 \times 10^{-6}$ & 320\tablenotemark{g}\\
SN\,1986O & 1987 Feb 12 & C/D & 1.4 & 0.80 & 72 & 69.79 & $1.6 \times 10^{-6}$ & 500\\ 
\nodata & 1987 Feb 12 & C/D & 4.8 & 0.07 & 72 & 20.94 & $1.0 \times 10^{-6}$ & 400\\ 
\nodata & 1987 Feb 12 & C/D & 15.0 & 0.18 & 72 & 168.24 & $7.9 \times 10^{-6}$ & 4000\\ 
\nodata & 1987 Apr 11 & D & 4.8 & 0.08 & 130 & 23.93 & $2.0 \times 10^{-6}$ & 250\\ 
\nodata & 1987 Apr 11 & D & 15.0 & 0.16 & 130 & 149.55 & $1.0 \times 10^{-5}$ & 2500\tablenotemark{g}\\ 
\nodata & 1987 May 24 & D & 4.8 & 0.08 & 173 & 23.93 & $2.5 \times 10^{-6}$ & 200\\ 
\nodata & 1987 May 24 & D & 15.0 & 0.15 & 173 & 140.20 & $1.3 \times 10^{-5}$ & 2000\tablenotemark{g}\\ 
\nodata & 1987 Aug 28 & A & 4.8 & 0.07 & 269 & 20.94 & $3.2 \times 10^{-6}$ & 160\\ 
SN\,1987D & 1987 May 15 & D & 4.8 & 0.20 & 45 & 49.13 & $1.3 \times 10^{-6}$ & 1000\\ 
\nodata & 1987 Jun 4 & D & 4.8 & 0.26 & 65 & 63.87 & $2.0 \times 10^{-6}$ & 1000\\ 
\nodata & 1987 Jun 21 & A & 4.8 & 0.06 & 82 & 14.74 & $1.0 \times 10^{-6}$ & 250\\ 
\nodata & 1987 Sep 18 & A & 4.8 & 0.07 & 171 & 17.20 & $2.0 \times 10^{-6}$ & 160\\ 
SN\,1987N & 1987 Dec 20 & B & 1.4 & 0.13 & 18 & 12.06 & N/A & N/A\\  
\nodata & 1987 Dec 20 & B & 4.8 & 0.10 & 18 & 31.81 & $5.0 \times 10^{-7}$ & 1600\\  
\nodata & 1988 Jan 12 & B & 1.4 & 0.10 & 41 & 9.28 & $2.0 \times 10^{-7}$ & 130 \\  
\nodata & 1988 Jan 12 & B & 4.8 & 0.08 & 41 & 25.45 & $7.9 \times 10^{-7}$ & 630\\  
\nodata & 1988 Feb 1 & B & 1.4 & 0.11 & 61 & 10.21 & $2.5 \times 10^{-7}$ & 100\\  
\nodata & 1988 Feb 1 & B & 4.8 & 0.09 & 61 & 28.63 & $1.0 \times 10^{-6}$ & 500\\  
\nodata & 1988 Mar 31 & C & 1.4 & 0.23 & 120 & 21.34 & $7.9 \times 10^{-7}$ & 100\\  
\nodata & 1988 Mar 31 & C & 4.8 & 0.07 & 120 & 22.27 & $1.6 \times 10^{-6}$ & 250\\  
\nodata & 1988 Apr 11 & C & 1.4 & 0.22 & 131 & 20.41 & $7.9 \times 10^{-7}$ & 100 \\  
\nodata & 1988 Apr 11 & C & 4.8 & 0.40 & 131 & 127.23 & $5.0 \times 10^{-6}$ & 1000\\  
\nodata & 1988 Aug 22 & D & 4.8 & 0.08 & 264 & 25.45 & $3.2 \times 10^{-6}$ & 160\\  
SN\,1989B & 1989 Feb 2 & A & 4.8 & 0.08 & 12 & 2.34 & $5.0 \times 10^{-8}$ & 200\\  
\nodata & 1989 Feb 2 & A & 8.4 & 0.06 & 12 & 3.07 & $7.9 \times 10^{-8}$ & 400\\  
\nodata & 1989 Feb 3 & A & 4.8 & 0.03 & 13 & 0.88 & $2.5 \times 10^{-8}$ & 79\\  
\nodata & 1989 Feb 3 & A & 8.4 & 0.03 & 13 & 1.53 & $5.0 \times 10^{-8}$ & 250\\  
\nodata & 1989 Mar 6 & A/B & 4.8 & 0.06 & 44 & 1.75 & $1.6 \times 10^{-7}$ & 79\\  
\nodata & 1989 Mar 27 & B & 4.8 & 0.06 & 65 & 1.75 & $2.0 \times 10^{-7}$ & 63\\  
\nodata & 1989 Apr 6 & B & 4.8 & 0.07 & 75 & 2.04 & $2.5 \times 10^{-7}$ & 63\\  
\nodata & 1989 May 15 & BnC & 4.8 & 0.06 & 114 & 1.75 & $3.2 \times 10^{-7}$ & 40\\  
SN\,1989M & 1989 Jul 17 & C & 4.8 & 0.13 & 33 & 15.19 & $5.0 \times 10^{-7}$ & 500\\ 
\nodata & 1989 Sep 4 & C & 4.8 & 0.10 & 82 & 11.68 & $7.9 \times 10^{-7}$ & 200\\ 
\nodata & 1989 Oct 24 & C/C & 4.8 & 0.07 & 132 & 8.18 & $1.0 \times 10^{-6}$ & 130\\ 
\nodata & 1989 Dec 21 & D & 4.8 & 0.05 & 190 & 5.84 & $1.0 \times 10^{-6}$ & 63\\ 
\nodata & 1990 Feb 13 & D/A & 4.8 & 0.08 & 244 & 9.35 & $1.6 \times 10^{-6}$ & 79\\ 
\nodata & 1990 May 29 & A & 4.8 & 0.15 & 349 & 17.53 & $3.2 \times 10^{-6}$ & 100\\ 
SN\,1990M & 1990 Jun 29 & A/B & 8.4 & 0.04 & 40 & 8.18 & $5.0 \times 10^{-7}$ & 400\\ 
\nodata & 1990 Dec 14 & C & 8.4 & 0.04 & 208 & 8.18 & $2.0 \times 10^{-6}$ & 130\\ 
SN\,1991T & 1991 May 8 & D & 8.4 & 0.05 & 30 & 5.01 & $3.2 \times 10^{-7}$ & 320\\
\nodata & 1991 May 8 & D & 15.0 & 0.06 & 30 & 10.73 & $6.3 \times 10^{-7}$ & 1000\\
\nodata & 1991 Jul 9 & A & 8.4 & 0.06 & 92 & 6.01 & $7.9 \times 10^{-7}$ & 200\\
\nodata & 1991 Jul 9 & A & 15.0 & 0.20 & 92 & 35.78 & $3.2 \times 10^{-6}$ & 1000\\
SN\,1991bg & 1991 Dec 26 & B & 1.4 & 0.26 & 26 & 6.40 & $1.0 \times 10^{-7}$ & 130 \\    
\nodata & 1991 Dec 26 & B & 8.4 & 0.16 & 26 & 23.64 & $6.3 \times 10^{-7}$ & 1300 \\ 
SN\,1992A & 1992 Jan 27 & B/C & 4.8 & 0.03 & 25 & 2.84 & $1.3 \times 10^{-7}$ & 160\\ 
\nodata & 1992 Oct 9 & A & 4.8 & 0.15 & 281 & 14.20 & $2.5 \times 10^{-6}$ & 100\\ 
SN\,1994D & 1994 May 4 & A & 4.8 & 0.06 & 63 & 3.94 & $3.2 \times 10^{-7}$ & 130\\
SN\,1995al & 1995 Nov 8 & B & 1.4 & 0.08 & 17 & 4.97 & $5.0 \times 10^{-8}$ & 130\\ 
SN\,1996X & 1996 May 29 & DnC & 8.4 & 0.09 & 59 & 28.75 & $1.6 \times 10^{-6}$ & 790\\
SN\,1998bu & 1998 May 13 & A & 8.4 & 0.07 & 12 & 4.33 & $1.0 \times 10^{-7}$ & 630\\
\nodata & 1998 May 31 & A & 8.4 & 0.04 & 30 & 2.47 & $2.0 \times 10^{-7}$ & 200\\
\nodata & 1998 May 31 & A & 15.0 & 0.18 & 30 & 19.88 & $1.0 \times 10^{-6}$ & 1600\\
\nodata & 1998 Jun 9 & BnA & 8.4 & 0.05 & 39 & 3.09 & $2.5 \times 10^{-7}$ & 200 \\
\nodata & 1998 Jun 9 & BnA & 15.0 & 0.23 & 39 & 25.40 & $1.3 \times 10^{-6}$ & 1600\\	
\nodata & 1999 Jan 7 & C & 4.8 & 0.07 & 251 & 2.47 & $7.9 \times 10^{-7}$ & 32\\
\nodata & 1999 Jan 7 & C & 8.4 & 0.06 & 251 & 3.71 & $1.3 \times 10^{-6}$ & 63\\
SN\,1999by & 1999 May 7 & D & 8.4 & 0.07 & 10 & 11.59 & $1.6 \times 10^{-7}$ & 1600\\    
\nodata & 1999 May 24 & D & 8.4 & 0.08 & 27 & 13.25 & $5.0 \times 10^{-7}$ & 790\\  
SN\,2002bo & 2002 May 21 & BnA & 8.4 & 0.06 & 75 & 19.17 & $1.3 \times 10^{-6}$ & 500\\  
\nodata & 2002 May 21 & BnA & 15.0 & 0.28 & 75 & 159.74 & $7.9 \times 10^{-6}$ & 4000\\  
\nodata & 2002 May 21 & BnA & 22.0 & 0.36 & 75 & 301.21 & $1.3 \times 10^{-5}$ & 8000\\  
\nodata & 2002 May 21 & BnA & 43.0 & 0.62 & 75 & 1013.94 & $4.0 \times 10^{-5}$ & 40,000\tablenotemark{g}\\  
\nodata & 2002 Jun 12 & B & 1.4 & 0.06 & 97 & 3.19 & $2.0 \times 10^{-7}$ & 25\\  
\nodata & 2002 Jun 12 & B & 4.8 & 0.08 & 97 & 14.60 & $1.0 \times 10^{-6}$ & 250\\  
\nodata & 2002 Jun 12 & B & 8.4 & 0.07 & 97 & 22.36 & $2.0 \times 10^{-6}$ & 500\\  
SN\,2002cv & 2002 May 21 & BnA & 8.4 & 0.06 & 17 & 19.17 & $4.0 \times 10^{-7}$ & 1600\\   		
\nodata & 2002 May 21 & BnA & 15.0 & 0.28 & 17 & 159.74 & $2.5 \times 10^{-6}$ & 13,000\\   		
\nodata & 2002 May 21 & BnA & 22.0 & 0.36 & 17 & 301.21 & $4.0 \times 10^{-6}$ & 25,000\\   		
\nodata & 2002 May 21 & BnA & 43.0 & 0.62 & 17 & 1013.94 & $1.3 \times 10^{-5}$ & 100,000\\   		
\nodata & 2002 Jun 12 & B & 1.4 & 0.06 & 39 & 3.19 & $7.9 \times 10^{-8}$ & 50 \\   		
\nodata & 2002 Jun 12 & B & 4.8 & 0.08 & 39 & 14.60 & $5.0 \times 10^{-7}$ & 400\\   		
\nodata & 2002 Jun 12 & B & 8.4 & 0.07 & 39 & 22.36 & $1.0 \times 10^{-6}$ & 1000\\   	
SN\,2003hv & 2003 Oct 21 & B & 8.4 & 0.05 & 59 & 9.23 & $7.9 \times 10^{-7}$ & 320\\
SN\,2003if & 2003 Oct 21 & B & 8.4 & 0.05 & 68 & 10.22 & $7.9 \times 10^{-7}$ & 320\\
PTF\,10iuv\tablenotemark{b}  & 2010 Aug 25.1 & D & 8.5 & 0.063 & 88 & 312.96 & $1.0 \times 10^{-5}$ & 4000 \\
\nodata\tablenotemark{b}  & 2011 May 12.2 & BnA & 8.5  & 0.032 & 348 & 158.96 & $2.0 \times 10^{-5}$ \tablenotemark{g}  & 790\tablenotemark{g}\\

PTF\,11kx\tablenotemark{c} & 2011 Mar 30.2 & B & 8.4 & 0.023 & 90 & 433.39 & $1.3 \times 10^{-5}$ & 5000\\

SN\,2011fe\tablenotemark{d} & 2011 Aug 25.0 & A & 8.5 &0.025 & 1 & 0.37 & $1.0 \times 10^{-9}$ & 158 \\

SN\,2014J\tablenotemark{e} & 2014 Jan 24.4 & B$\rightarrow$BnA & 22.0 & 0.008 & 7 & 0.03 & $4.0 \times 10^{-9}$ & 20 \\

iPTF\,14atg\tablenotemark{f} & 2014 May 16.2 & A & 6.1 & 0.010 & 13 & 28.27 & $4.0 \times 10^{-7}$ & 2000 \\
\nodata\tablenotemark{f} & 2014 May 16.2 & A & 22.0 & 0.010 & 13 & 103.67 & $1.6 \times 10^{-6}$ & 13,000\\

\enddata
\tablenotetext{a}{Assuming $\epsilon_B = 0.1$, $M_{\rm ej,Ch}=1$, and $E_{\rm K,51}=1$.}
\tablenotetext{}{All limits are from \cite{Panagia_etal06} except those marked with: $^{\rm b}$from \cite{Kasliwal_etal12}; $^{\rm c}$\cite{Dilday_etal12}; $^{\rm d}$\cite{Horesh_etal11}; $^{\rm e}$\cite{Chandler_Marvil14} (see also \citealt{Perez-Torres_etal14}); and $^{\rm f}$\cite{Cao_etal15}.}
\tablenotetext{g}{This limit is not sufficiently constraining to be completely described by the model of outer SN ejecta applied in this study. This radio limit is consistent with dense CSM which leads to rapid deceleration of the SN blast. It is therefore possible that the steep outer ejecta are not interacting with the CSM at the time of this observation, but rather the inner shallower ejecta are interacting.}

\end{deluxetable}

\end{document}